\documentclass[%
 reprint,
superscriptaddress,
showkeys,
 amsmath,amssymb,
 aps,
 prx,
 floatfix,
]{revtex4-2}

\usepackage{graphicx}
\graphicspath{{figures/}}
\usepackage[export]{adjustbox}  
\usepackage{chngcntr}  
\usepackage{dcolumn}
\usepackage{bm}
\usepackage[colorlinks=true, allcolors=blue]{hyperref}
\usepackage{siunitx}
\usepackage{upgreek}
\usepackage{makecell}
\usepackage[misc,clock,geometry,electronic]{ifsym}

\usepackage{xcolor}
\definecolor{darkgreen}{rgb}{0.1,0.6,0.1}
\definecolor{steelblue}{rgb}{.273,.508,.703}

\newcommand{\Potassium}{$^{40}$K}
\newcommand{\Thorium}{$^{232}$Th}
\newcommand{\Uranium}{$^{238}$U}

\begin{document}

\title{Spectroscopic measurements and models of energy deposition in the substrate of quantum circuits by natural ionizing radiation}

\newcommand{\affilQSD}{\affiliation{Quantum Sensors Division, National Institute of Standards and Technology, 325 Broadway, Boulder, Colorado USA 80305}}
\newcommand{\affilCU}{\affiliation{Department of Physics, University of Colorado, Boulder, Colorado USA 80309}}
\newcommand{\affilRPD}{\affiliation{Radiation Physics Division, National Institute of Standards and Technology, 100 Bureau Drive, Gaithersburg, Maryland USA 20899}}
\newcommand{\affilPNNL}{\affiliation{Pacific Northwest National Laboratory, Richland, Washington USA 99354}}

\author{Joseph W. Fowler}\altaffiliation[Corresponding author: ]{\href{mailto:Joe Fowler <joe.fowler@nist.gov>}{Joe.Fowler@nist.gov}}\affilQSD\affilCU
\author{Paul Szypryt}\affilQSD\affilCU
\author{Raymond Bunker}\affilPNNL
\author{Ellen R. Edwards}\affilPNNL
\author{Ian Fogarty Florang}\affilCU\affilQSD
\author{Jiansong Gao}\altaffiliation[Currently at ]{Amazon Web Services Center for Quantum Computing, Pasadena, California USA}\affilQSD
\author{Andrea Giachero}\affilCU\affilQSD\affiliation{Department of Physics, University of Milano-Bicocca, Milan, Italy}
\author{Shannon F. Hoogerheide}\affilRPD
\author{Ben Loer}\affilPNNL
\author{H. Pieter Mumm}\affilRPD
\author{Nathan Nakamura}\affilQSD\affilCU
\author{Galen C. O'Neil}\affilQSD
\author{John L. Orrell}\affilPNNL
\author{Elizabeth M. Scott}\affiliation{Department of Physics, Centre College, 600 West Walnut Street, Danville, Kentucky USA 40422}
\author{Jason Stevens}\affilQSD\affilCU
\author{Daniel S. Swetz}\affilQSD
\author{Brent A. VanDevender}\affilPNNL
\author{Michael Vissers}\affilQSD
\author{Joel N. Ullom}\affilQSD\affilCU

\date{\today}

\begin{abstract}
Naturally occurring background radiation is a potential source of correlated decoherence events in superconducting qubits that will challenge error-correction schemes.  In order to characterize the radiation environment in an unshielded laboratory representative of superconducting qubits' environments, we performed broadband, spectroscopic measurements of background radiation events inside a millikelvin refrigerator. The spectrometer was designed to mimic the size and composition of a quantum circuit.  Specifically, we measured the background radiation spectra in silicon substrates of two thicknesses, \qty{500}{\micro\m} and \qty{1500}{\micro\m}, and one area, \qty{25}{mm^2}. The observed spectra span energies from a few keV up to nearly \qty{10}{MeV}, are nearly featureless, and decrease in intensity by a factor of \qty{40000}{} between \qty{100}{keV} and \qty{3}{MeV} for the \qty{500}{\micro\m} substrate. We integrate the spectra to obtain the average event rates and deposited power levels. These quantities correspond to a rate of 0.023 events per second and a power of \qty{4.9}{keV.s^{-1}}, when counting events that deposit at least \qty{40}{keV} for the \qty{500}{\micro\m}-thick substrate.
We find the cryogenic measurements are in good agreement with predictions based on simple measurements of the terrestrial gamma-ray flux outside the refrigerator, published models of cosmic-ray fluxes, a crude model of the cryostat, and radiation-transport simulations.  This model requires no free parameters to predict the background radiation spectra in the silicon substrates. The agreement between measurements and predictions demonstrates that the model we present can be used to assess the relative contributions of terrestrial and cosmic-ray sources to background radiation interactions in silicon substrates of varying thickness. These spectroscopic measurements are performed with a novel combination of superconducting microresonators located on micromachined silicon islands that define the interaction volume with background radiation.  The resonators transduce deposited energy to a readily detectable electrical signal. Microresonator readout closely resembles dispersive superconducting qubit readout, so similar devices---{with or without micromachined islands}---are suitable for integration with superconducting quantum circuits as detectors for background radiation events. For our specific laboratory conditions, we find that gamma-ray emissions from radioisotopes are responsible for the majority of events that deposit $E<$\qty{1}{MeV}.
We present results demonstrating that the background radiation spectrum contains relevant contributions from cosmic-ray particles other than muons, particularly a tail of multi-MeV events due to protons and neutrons.
These observations suggest several paths to reducing the impact of background radiation on quantum circuits, supported by an empirically validated model for generating reliable predictions of radiation interactions with silicon substrates.
 
\end{abstract}

\keywords{Superconducting devices and qubits; Interactions of radiation with matter; Thermal kinetic inductance detectors}

\maketitle

\section{Introduction}
\label{sec:intro}

There is widespread interest in developing quantum computers that use qubits based on a variety of physical systems, including systems in which the qubits are deposited on or in silicon substrates.  Here we focus on superconducting qubits~\cite{Bravyi2022, Kjaergaard2020}, although some spin-based qubits also use silicon substrates~\cite{Burkard2023}.  Improving the coherence of individual qubits is critical to realizing a practical quantum computer. Naturally occurring background radiation is a potential source of decoherence that will be present even in ideal qubits that are perfect in design and manufacture. Sources of background radiation include energetic charged particles and gamma rays created by the interaction of cosmic rays with the Earth's atmosphere. The decay of ubiquitous terrestrial radioisotopes is another source, producing gamma rays as well as less penetrating alpha and beta particles. Techniques have been developed to reduce radiation backgrounds, but these are often extreme in nature.  They include conducting experiments deep underground to attenuate cosmic radiation sources, using massive shields to attenuate terrestrial gamma rays, and constructing experiments exclusively from materials with high radiopurity.  Given the cost and inconvenience of implementing these techniques, it is likely that commercially viable quantum computers must be robust against the radiation backgrounds found in conventional, unshielded laboratories.  As a result, it is important to understand the intensity and spectral distribution of background radiation events that future quantum circuits will encounter.  


Naturally occurring ionizing radiation is known to disrupt superconducting circuits. It has been suggested as a source of the excess quasiparticle population observed in qubits~\cite{Vepsalainen2020,McEwen2022} and in other superconducting devices~\cite{Zmuidzinas2012}.  The same work showed that ionizing radiation limited the qubits' coherence time to a few milliseconds, while use of a gamma-ray shield produced a small improvement.  Ionizing radiation has also been shown to produce transient events in superconducting microresonators~\cite{Cardani2021}. Operating the same devices underground to shield against cosmic rays reduced the transient-event rate by a factor of 30 and improved the quality factor of the resonators by a factor of four.  Other authors~\cite{Wilen2021, McEwen2022,Thorbeck2023} operating several qubits on the same substrate have observed ``error bursts,'' in which the coherence of multiple qubits is simultaneously disrupted, even for qubits separated by several millimeters on their shared substrate.  These error bursts reveal both the danger to error-correction schemes and the likelihood of the root cause being energy deposition in the common substrate. Various techniques have been proposed or demonstrated to reduce the energy that propagates into a thin-film circuit from radiation interactions in the supporting substrate~\cite{Karatsu2019, Martinis2021, Iaia2022}.  Despite this growing body of literature, which includes simulations of the expected spectrum of substrate events due to background radiation~\cite{Wilen2021, Cardani2021}, this spectrum has not previously been measured.  Details of the disruptive events are important for the development of successful schemes to protect qubits from background radiation and for understanding the limits of such efforts.

A superconducting quantum circuit typically consists of metal and dielectric thin films deposited and patterned on a piece of silicon with planar dimensions near \qty{1}{cm^2}.  The films for a single qubit often occupy only a small fraction of the area of the substrate, and their thickness is at most hundreds of nanometers, while the substrate is generally a thousand times thicker. Incident radiation is thus far more likely to interact with the comparatively massive substrate than it is to interact directly with the circuit's thin films.  Energy deposited in the substrate will propagate as phonons, and to a lesser extent as energetic charge carriers. Some of these excitations will interact with the films on the substrate surface, generating quasiparticles in a superconductor, causing decoherence in a qubit, or otherwise degrading circuit performance~\cite{Vepsalainen2020,McEwen2022,Martinis2021}.
The propagation of energy from the substrate into the surface films and the effects on circuit performance are also important. Because these consequences depend strongly on the details of the circuit design, they are not discussed here. Instead, this report focuses on the dominant interaction of background radiation events with silicon substrates.


While extensive prior work has characterized natural radiation backgrounds, the topic is sufficiently broad and complex that a simplified and focused description of likely radiation backgrounds is not yet available to the quantum information science (QIS) community.  Radiation described as ``cosmic rays'' has contributions from muons, electrons, protons, neutrons, and gamma rays. Several cosmic-ray models are available. Their predictions vary somewhat, and fluxes depend strongly on altitude and weakly on additional factors such as weather, geomagnetic latitude, and solar activity.  Radiation from terrestrial radioisotopes depends on the local geology and the composition of nearby construction.  Predicting the effects of external radiation on a circuit substrate also requires particle-transport calculations that track the type of radiation and model the shielding effects of buildings and cryostats.  Hence, one contribution of this work is a clear description of how we characterized the radiation background in our laboratory so that such a methodology could be adapted for the readers' own purposes.

\begin{figure*}[thb]
    \centering
    \includegraphics[width=\textwidth,frame]{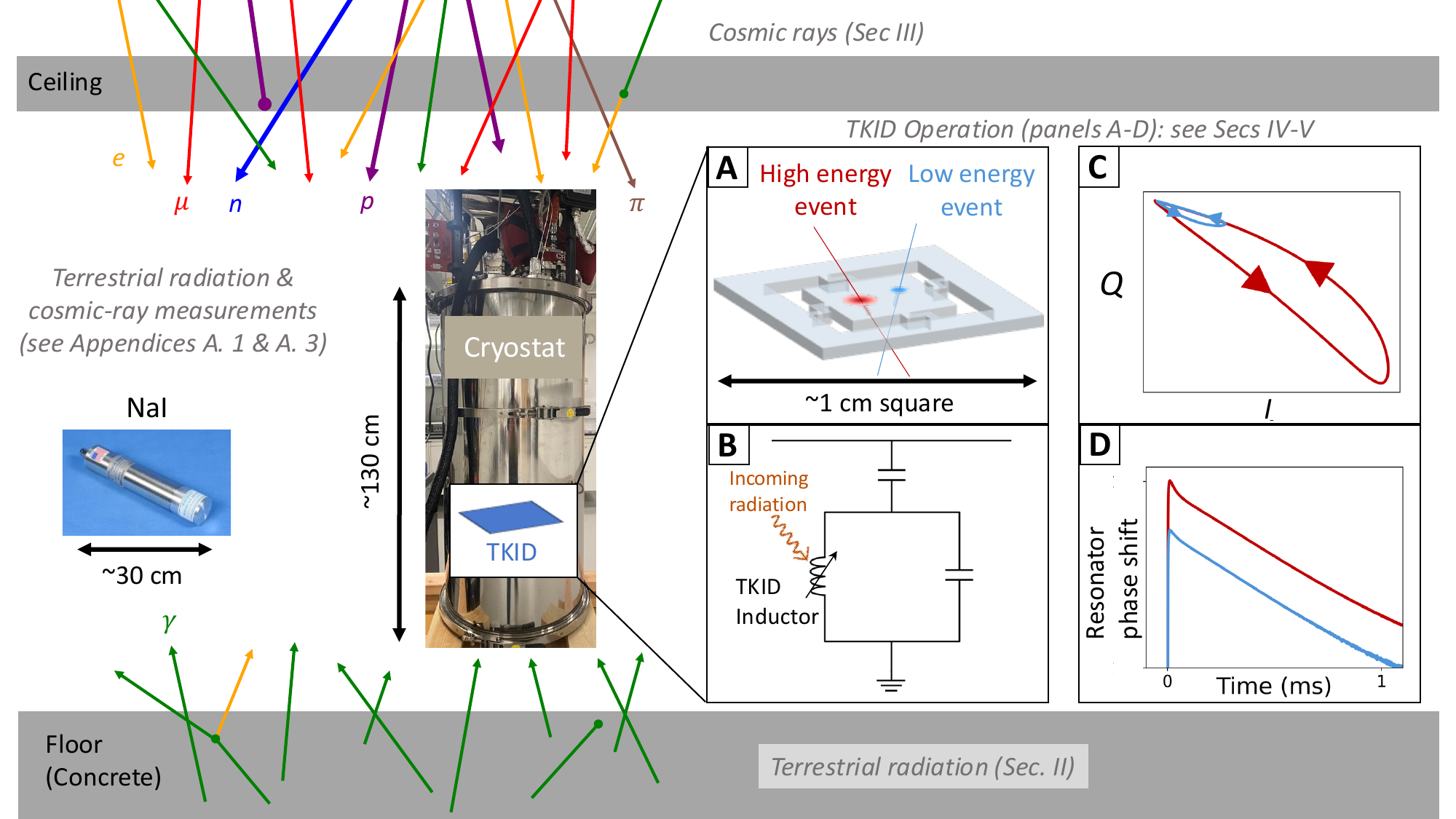}
    \caption{Overview of the measurements made in this work. A superconducting spectroscopic sensor on a silicon substrate is operated in a millikelvin cryostat (center). The sensor (inset \textbf{A}) is a Thermal Kinetic-Inductance Detector (TKID), a radiation-sensitive inductor embedded in a resonant circuit (\textbf{B}). Cosmic rays of many species arrive from above, partially screened by the concrete ceiling. Radioisotopes in the floor emit gamma rays and Compton-scattered electrons. Heating of the TKID by absorbed radiation causes transient changes in the resonator's complex RF transmission (\textbf{C}). The changes are characterized by an angle in the complex plane. They have a rapid onset and return to the resting state in an exponential ``pulse'' (\textbf{D}). A conventional NaI detector (left) is also used to measure the intensity of both cosmic and terrestrial radiation sources outside the cryostat. Given the intensity measured in a conventional detector, the spectrum of energy deposited in the silicon TKID can be modeled. Aspects of the figure are annotated to identify where more descriptive detail is found within this report.}
    \label{fig:cartoon}
\end{figure*}

The primary contribution of this work is the validation of a background radiation model through spectroscopic \textit{measurements} performed directly in substrates that are representative of quantum circuits.
These measurements rely on a novel superconducting circuit placed on a semi-isolated silicon substrate (Figure~\ref{fig:cartoon}). This Thermal Kinetic-Inductance Detector (TKID) measures the energy deposited by individual background radiation events in a well-defined volume of the substrate, over a broad energy range. The TKID's design is intended to mimic that of a quantum circuit, enabling direct comparison between models and spectral measurements.

Meaningful comparative validation of such models has been limited or absent to date, because few energy-resolving radiation detectors are both sensitive over the relevant, broad energy range and similar in size and composition to quantum circuits.  The distribution of energy deposited by background radiation events depends strongly on the size of the interaction volume: the spectrum measured by a conventional, cylindrical radiation detector with dimensions of $\sim$\qty{100}{mm} differs substantially from the spectrum in a silicon wafer less than \qty{1}{mm} thick. In this report we describe an approach that lets us predict the small-device response from measurements made by traditional, large-volume radiation detectors, which we then confirm with TKID observations. We present both models and measurements of the background event rate, integrated power, and event spectrum as a function of substrate dimensions.

The following four sections of this report describe our modeling and instrumentation, as annotated in Figure~\ref{fig:cartoon}. In Section~\ref{sec:terrestrial}, we describe our model of terrestrial radiation backgrounds caused by radioactive decays that generate gamma rays. Measurements with a conventional scintillator spectrometer let us estimate the absolute activities of gamma emitters and thus predict the spectrum of energy deposition in millimeter-scale silicon devices such as our TKIDs or quantum circuits. Section~\ref{sec:CR} describes our model of secondary cosmic rays. As with the terrestrial sources, we measure the cosmic-ray spectrum with a conventional spectrometer and use the results to predict the spectrum of energy deposition in small silicon substrates. Section~\ref{sec:TKIDs} describes our TKID calorimeter, made of superconducting thin films on a silicon substrate, while Section~\ref{sec:measurements} describes the measurements of the radiation backgrounds made with TKIDs of two sizes.

Section~\ref{sec:comparison} compares the spectra measured with the TKIDs to predictions of the background radiation interactions with the TKIDs. Owing to the event rates determined by the conventional spectrometer, the predictions for the spectra in silicon substrates have no free parameters. The TKID spectra, therefore, confirm the models and allow us to predict background spectra over a wide range of substrate thicknesses (Section~\ref{sec:mitigation}). We outline possible strategies for mitigation of the effects of background radiation on quantum circuits.

This work provides a novel measurement of the spectrum of background radiation events in silicon substrates like those used in many quantum circuits. Our measurement technique, the use of cryogenic sensors based on microresonators, is suitable for wide adoption by the QIS community. Further, we have developed predictions for the background radiation spectrum that are informed by separate measurements of the local radiation fields outside our cryostat. These predictions and our measurements agree. 

\section{Terrestrial Radiation Backgrounds}
\label{sec:terrestrial}

The primary sources of ionizing radiation that penetrate a cryostat in a typical laboratory are the various species of secondary cosmic rays and the gamma radiation that accompanies the decay of common radioactive isotopes. We call these the cosmic-ray and terrestrial radiation backgrounds, respectively. The terrestrial background creates the majority of the interaction events and of the total energy deposited in millimeter-scale silicon devices in a typical laboratory environment, while cosmic rays are responsible for the rarer events that deposit the largest energies. We model the terrestrial radiation in this section and the cosmic rays in Section~\ref{sec:CR}.

For the purposes of modeling qubit radiation backgrounds, we use two silicon substrates \qty{5}{mm} square, \qtylist{500;1500}{\um} thick, with a thin superconducting circuit fabricated on one surface of each. These sizes correspond to the specific silicon microcalorimeters we used for the present measurements (Section~\ref{sec:TKIDs}), and more generally to the size, shape, and composition of a typical superconducting qubit. While few qubits might use substrates as thick as \qty{1500}{\um}, this size allows us to check how backgrounds scale with wafer thickness.

\subsection{Three predominant radioactive decay chains}

Trace levels of naturally occurring radioactive elements are present in a typical laboratory, residing in structural materials such as concrete and the underlying soil and rock~\cite{Papastefanou2005,Nemlioglu2022}. The gamma-ray spectrum emitted by these natural radiation sources consists of hundreds of distinct emission energies. The most common emitters include the primordial isotopes \Potassium, \Thorium, and \Uranium, whose lifetimes are all long enough to have survived from the time of the Earth's formation. \Potassium\ decays with \qty{11}{\percent} probability to $^{40}$Ar by electron capture, emitting a \qty{1.461}{MeV} gamma ray. Thorium and uranium give rise to two lengthy decay chains. The thorium chain includes twelve distinct decays, accompanied by both x-ray and gamma-ray emissions with energies as high as \qty{2.615}{MeV}. We model this chain by the 289 emissions with energies of at least \qty{50}{keV} and emitted by at least \qty{0.01}{\percent} of the decays of either thorium or its progeny. The uranium chain includes the decays of eighteen radioisotopes, yielding 246 gamma-ray and 33 x-ray emissions above \qty{50}{keV} and \qty{0.01}{\percent} probability.~\footnote{S. Chu, L. Ekstr{\"o}m, and R. Firestone, \href{http://nucleardata.nuclear.lu.se/toi/}{WWW Table of Radioactive Isotopes, database version 1999-02-28}.}
The most intense of these gamma-ray and x-ray emissions appear as distinctive ``lines'' or ``peaks'' in the response spectrum of typical gamma-ray spectrometers.

The spectrum incident on a quantum circuit, however, is not composed only of these distinct lines. A prominent continuum is also present, the result of gamma rays that have undergone Compton scattering before reaching or within the substrate~\footnote{Chapter 10 of G.F. Knoll's \textit{Radiation Detection and Measurement} (4th Edition, Wiley, Hoboken, 2010) provides an excellent and intuitive introduction to all the typical gamma-ray interaction processes occurring in ``macroscopic'' radiation detectors.}. To model these energy-loss mechanisms and the ejected energetic electrons, we used the GEANT4 particle-transport Monte Carlo code~\cite{Geant2003,Geant2006,Geant2016}, as driven by the Tool for Particle Simulation (TOPAS)~\cite{Perl2012,Faddegon2020}. We assume that the emitters are uniformly distributed throughout a slab of concrete \qty{22}{cm} thick. We treat this as an approximation for any realistic foundation under a laboratory building and for the underlying rock. The attenuation length in concrete for all relevant gamma rays (that is, energies no higher than \qty{2.615}{MeV}) is \qty{11}{cm} or less, so we consider a \qty{22}{cm}-thick distribution to be a useful proxy for the full thickness of the Earth's crust. This modeling approximation is thin enough to permit efficient simulations, yet thick enough to capture the energy distribution of photons emitted at a realistic range of depths in a dense material.

We fit models of the complex terrestrial gamma spectrum to measurements with only five intensity parameters. To achieve this simplification, we assume secular equilibrium among the isotopes within a given decay chain. That is, we take the relative proportions of isotopes in a chain to be constant, the result of an equilibrium between production and decay of each radioisotope but the first. Secular equilibrium is anticipated in natural materials, in the absence of chemical processing. One potential exception arises from the gaseous nature of radon: it could be enhanced or depleted in a material by mechanical means alone, so its production and decay need not be locally in equilibrium. Thus, we model the terrestrial background with one \Potassium\ intensity, plus potentially distinct intensities of the ``pre-radon'' and ``post-radon'' segments of the \Thorium\ and \Uranium\ decay chains: a total of five parameters. In practice, our measurements set no meaningful constraints on the pre-radon segment of the uranium chain.

The Supplemental Material (Section~\ref{sec:NaI_terrestrial}) describes measurements of terrestrial radiation we made using a commercial NaI scintillation spectrometer. From these spectra, we determined the absolute levels of gamma-ray activity per kg of concrete floor in our lab. This determination allows us to model the background in a small, silicon substrate without unknown, free parameters.

\subsection{Terrestrial backgrounds in silicon substrates}

 \begin{figure}
    \centering
    \includegraphics[width=\columnwidth]{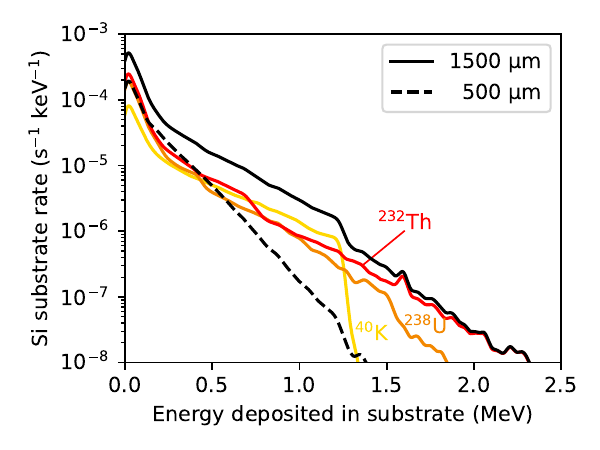}
    \caption{Predicted spectrum of energy deposited by terrestrial gamma-ray sources in a silicon substrate, \numproduct{5 x 5}{mm$^2$} square and \qty{1500}{\um} or \qty{500}{\um} thick. The three colored curves labeled by decay chain show the contributions of each to the spectrum for the thicker substrate.}
    \label{fig:gamma_qubit}
\end{figure}

Given the detector-independent activity values, we used TOPAS and GEANT4 to model the spectrum of energy that would be deposited in the \qtyproduct{5 x 5}{mm} silicon substrate of a quantum circuit after propagation through the cryostat. Figure~\ref{fig:gamma_qubit} shows the spectrum predicted for two substrate thicknesses (\qtylist{500;1500}{\um}), as well as the breakdown of one spectrum into the contributions from the different terrestrial sources. These predictions assume the activity level found in our main laboratory; the measurements made in several other locations demonstrate that the overall level can vary by factors of a few from one laboratory to another.  

While we model the complete spectrum, and eventually compare it to measurement (Section~\ref{sec:comparison}), we also find the rate of interactions and the rate of energy deposition in the substrate to be efficient summaries of the spectrum. The predicted interaction rates for terrestrial gamma rays to deposit more than \qty{40}{keV} in the silicon substrates are 0.0153 and 0.038 events per second for wafers of thickness \qtylist{500;1500}{\um}, respectively. These rates equate to mean time between events of \qty{65}{s} and \qty{26}{s}. Energy is deposited at corresponding rates of \qtylist{2.98;8.3}{keV s^{-1}}. These values have relative systematic uncertainties of $\pm$\qty{5}{\percent} due to limitations of the simplified model. The statistical uncertainties are smaller, $\pm$\qty{1}{\percent}. The cutoff of \qty{40}{keV} is chosen because the imperfect thermal isolation of our devices confuses measurement at lower energies (see Section~\ref{sec:measurements}). 

The rates of energy deposition for the \qty{1500}{\um} and \qty{500}{\um} substrate thicknesses, counting only events that leave at least \qty{40}{keV} in the substrate, are in a ratio of 2.8 to 1. Without the \qty{40}{keV} cutoff, the ratio becomes 3.1 to 1. Thus the power deposited in the silicon is approximately proportional to the ratio of volumes, as expected. The predicted event rate is only 2.5 times larger in the thicker wafer, however. As discussed in Section~\ref{sec:mitigation}, the rate of gamma rays interacting directly in the silicon is indeed in the expected 3:1 ratio. Other interactions result from secondary electrons, which are created by Compton scattering of gamma rays in material around the silicon. These electrons interact readily with even very thin silicon substrates, in a rate proportional not to the silicon volume but to the (effective) area. Secondary electrons therefore increase the total interaction rate in substrates of any thickness by similar amounts and reduce the ratio.

\section{Cosmic-Ray Backgrounds}
\label{sec:CR}

\subsection{Modeling the cosmic rays}

Primary cosmic rays---for the most part protons or other fully ionized nuclei---enter the Earth's atmosphere and interact high in the stratosphere with the nucleus of an air molecule. These relativistic collisions produce a spray of nuclear material which can induce further collisions in a growing particle population known as a cosmic-ray air shower~\cite{Kampert2012,Knapp2003}. The secondary cosmic rays in the air shower that reach the ground consist of a diverse assortment of particle types. They constitute the cosmic-ray background observed in a laboratory and relevant to QIS instruments. The secondaries include muons and electrons of either charge, protons, neutrons, and gamma rays. Even short-lived, unstable particles such as pions appear in very small numbers. We find that terrestrial gamma rays exceed the cosmic-ray background in terms of event rate and power deposition in a silicon substrate in our laboratory, though it is important to model both background sources. The cosmic-ray protons and neutrons are responsible for essentially all events in which at least \qty{2}{MeV} is deposited.

Many models of the cosmic-ray spectrum are available in the literature. Some describe the spectra of all particle types, while others are restricted to muons, the most penetrating species. Some parameterize the results of air-shower simulations; others represent a purely empirical synthesis of cosmic-ray measurements. Wanting to model all particles (and not only muons), we use the PARMA~4.0 tool~\cite{Sato2016}, an analytical approximation of radiation in the atmosphere. PARMA parameterizes the distribution of energies and zenith angles expected for each of several particle types, at the elevation (or atmospheric depth) of the user's choice. The parameters are fitted to match the results of air-shower simulations performed with the Particle and Heavy Ion Transport System (PHITS)~\cite{Sato2013}. Further details of the cosmic-ray models are given in the Supplemental Material, Section~\ref{sec:CR_details}, while Section~\ref{sec:CR_NaI} describes our use of a conventional scintillation spectrometer to calibrate the cosmic-ray models.

The measurements described here were made in Boulder, Colorado, at \qty{1650}{m} above sea level. The cosmic-ray muon flux is approximately \qty{33}{\percent} higher in Boulder than at sea level, a modest and well-understood difference. Other species are more sensitive to elevation than muons are. For example, the proton flux in Boulder is some four times higher than at sea level.   Still, our results are generally applicable with the understanding that the relative amplitude of the nucleon-induced high-energy tail will depend on elevation of an observation.

\subsection{Cosmic rays in a qubit circuit}
\label{sec:CR_qubit}

\begin{figure}
    \centering
    \includegraphics[width=\columnwidth]{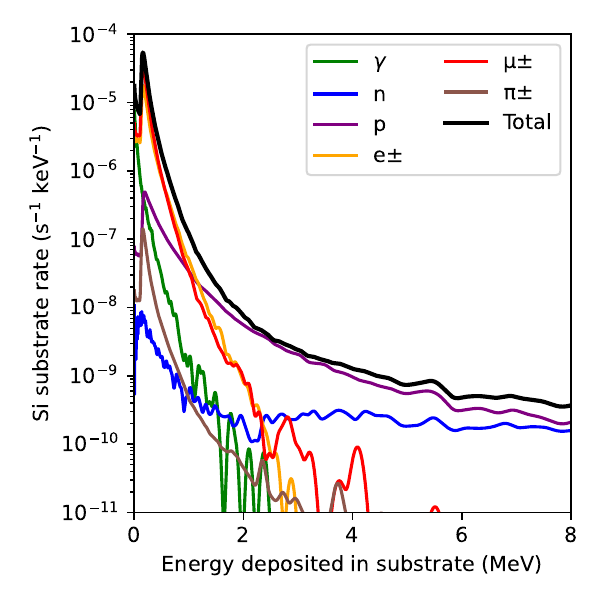}
    \caption{
    Energies deposited by cosmic rays (modeled) in a silicon substrate (\qty{500}{\um} thick, 25 mm$^2$ area), both in total (top curve) and separated by the species of the particle incident upon the circuit. The $\mu^\pm$ and $e^\pm$ dominate the interactions by numbers, but nucleons ($p$ and $n$) cause all of the rare, most energetic events. Charged particles (other than protons) are shown with both charge states combined. The spectra for both charge states of $\mu$ and $\pi$ are very similar, while $e^-$ outnumber $e^+$ by a 2:1 ratio over most of the spectrum.
    }
    \label{fig:CR_TKID_components}
\end{figure}

\begin{figure}[tbh]
    \centering
    \includegraphics[width=\columnwidth]{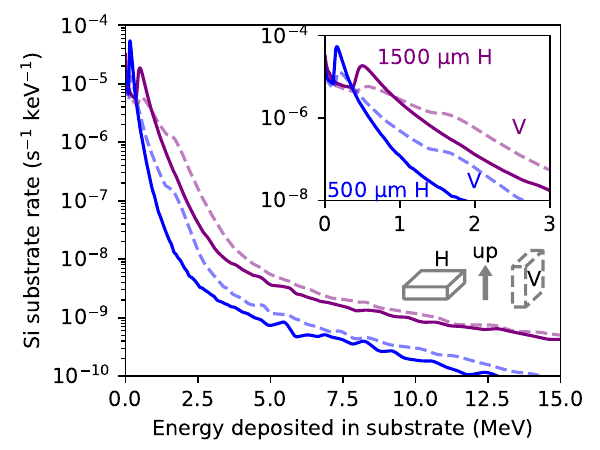}
    \caption{Simulated energy deposited by cosmic rays (summed over all particle species) in \qtyproduct{5 x 5}{mm} silicon substrates of \qty{1500}{\um} and \qty{500}{\um} thickness when oriented horizontally (solid curves) or vertically (dashed). The horizontal \qty{500}{\um} result is the same as the curve labeled ``Total'' in Figure~\ref{fig:CR_TKID_components}. Prominent peaks (at \qtylist{0.52;0.17}{MeV}) in the horizontal model correspond to the typical energy loss of minimum-ionizing $\mu^\pm$ and $e^\pm$ passing through silicon along paths approximately equal to the wafer thickness. The cartoon shows our definitions of horizontal (``H'') and vertical (``V'') orientation; the inset highlights the spectra at lower energies.}
    \label{fig:CR_TKID_loss}
\end{figure}

We model a qubit substrate as in Section~\ref{sec:terrestrial}: as a \qtyproduct{5 x5 }{mm} square of silicon, diced from a wafer with one of two thicknesses (again, \qtylist{500;1500}{\um}). Cosmic rays generated by the PARMA model must first pass through intervening materials---the concrete ceiling and aluminum cryostat shells---before entering the silicon substrate. At our higher-elevation site, the model predicts the rate of events depositing at least \qty{40}{keV} to be \qtylist{0.0075;0.0090}{s^{-1}} for the wafers of \qtylist{500;1500}{\um} thickness, respectively. The corresponding powers are \qtylist{1.91;5.3}{keV.s^{-1}}. The uncertainties in these models are dominated by systematic uncertainties that we estimate to be \qty{3}{\percent}; they include the effective volume of the NaI spectrometer used to calibrate the cosmic-ray intensity and the angular distribution and particle-species breakdown of cosmic rays.

The cosmic-ray event rate differs only slightly between the two silicon wafers of very different thickness, while the power deposited scales approximately as the volume. In contrast, for terrestrial gamma rays, both event rate and power scaled approximately with volume. The scaling rules differ because the charged particles that constitute most cosmic rays lose energy almost continuously in multiple scattering events per mm in silicon~\cite{Workman:2022ynf}, while a similar amount of silicon is optically thin to photoelectric absorption and Compton scattering at the relevant (MeV-scale) energies. Thus, an energy-loss event is nearly certain for cosmic rays that pass through the silicon wafer but unlikely for any single gamma ray.

Figure~\ref{fig:CR_TKID_components} shows the distribution of energy deposited by cosmic rays in a silicon substrate, both the overall spectrum and its breakdown by particle species. The distribution shows that the event rate and total power are dominated by charged leptons, primarily the $\mu^\pm$ but also, to a lesser extent, the $e^\pm$. On the other hand, the rare events that deposit more than \qty{2}{MeV} in a \qty{500}{\micro m}-thick silicon wafer are almost entirely the result of cosmic-ray protons and neutrons. This result suggests different mitigation strategies, depending on whether the more frequent leptonic events  or the more energetic but rarer hadronic events are more disruptive to a given circuit.

One tactic sometimes employed to mitigate cosmic-ray events is to orient a qubit circuit vertically (that is, to align one long dimension of the chip vertically)~\cite{loer2024}. Figure~\ref{fig:CR_TKID_loss} shows that this approach changes the spectrum only modestly. Although a vertical orientation does reduce the rate of low-energy events, it increases the rate of higher-energy events. Overall, a vertical orientation reduces the expected rate of cosmic-ray events by approximately one-third, but the mean energy deposited per event grows by a similar factor, so that the power deposited is reduced by only \qty{2}{\percent} for our specific geometry. Because gamma rays emitted isotropically from a large, unshielded floor are likely to be even less affected by circuit orientation, this particular mitigation tactic would appear to offer limited benefits.

Cosmic rays produce events at a wide range of energies, but at energies below $\sim$\qty{1.5}{MeV}, they contribute less to the predicted spectrum than the gamma rays emitted by terrestrial sources under the conditions found in our laboratories. Cosmic-ray secondaries are still important, however, because their energies can be much higher than the $\sim$\qty{2.6}{MeV} maximum of the gamma-ray events. Nucleons (protons and neutrons) in the cosmic rays are thus responsible for the upper region of the absorbed-energy spectrum (Figure~\ref{fig:CR_TKID_components} for the \qty{500}{\um}-thick silicon, or Supplemental Material Figure~\ref{fig:CR_TKID_thick_components} for the \qty{1500}{\um} substrate). Section~\ref{sec:mitigation} further discusses the relative importance of cosmic rays and terrestrial gamma rays for devices on silicon wafers, which depends both on the substrate thickness and the circuit's sensitivity to the lower-energy events.

\section{Thermal Kinetic Inductance Detectors as a Probe of Backgrounds}
\label{sec:TKIDs}

The results shown so far consist of simulated backgrounds in a silicon substrate typical of superconducting circuits. Measurements made in a large, conventional scintillator spectrometer allowed us to calibrate the intensities of these simulated backgrounds. We have also made measurements with a type of cryogenic, calorimetric detector very similar to a typical superconducting qubit in most respects. In this section, we describe that detector. It is uniquely suited to the goal of validating the background models, as it enables spectroscopic study of the energy deposited in a volume of silicon similar to the substrate of a quantum circuit, over a suitably wide range of energies.

The microcalorimeters used in this study are based on the microwave kinetic inductance detector (MKID)~\cite{Day2003, Zmuidzinas2012}. The MKID is a cryogenic detector technology that exploits the exceptionally high kinetic inductance of a superconductor to measure the energy of incident radiation events. In an MKID, a thin superconducting film is fabricated into a microresonator circuit. When energy is deposited in the superconducting resonator, Cooper pairs are broken. With fewer pairs to carry the supercurrent, the kinetic inductance grows and the resonant frequency falls. The magnitude of the frequency shift depends on the energy absorbed in the radiation interaction event, enabling spectroscopic measurements. 


A thermal kinetic inductance detector (TKID) calorimeter~\cite{Cecil2012, Cecil2015, Ulbricht2015} incorporates a structure to absorb and thermalize energy. The absorber can be co-located with the superconducting resonator and semi-isolated on a micromachined island to confine thermal energy long enough for accurate measurement.  In this work, no separate absorbing structure is used; the micromachined island itself is the radiation absorber.  The TKID is similar in many ways to the more mature transition-edge sensor (TES) calorimeter~\cite{Irwin1996, Irwin2005, Ullom2015}, but its silicon absorber and well-defined interaction volume made it especially suited for this work.

Being a less mature detector technology, however, TKIDs are not without their disadvantages. In particular, the device physics is less thoroughly understood, and the energy resolution of current devices for photon detection is significantly worse than that achieved in the best TESs. For the current measurement, the broad dynamic range is far more important than the energy resolution.

\begin{figure}
    \centering
    \includegraphics[width=\columnwidth]{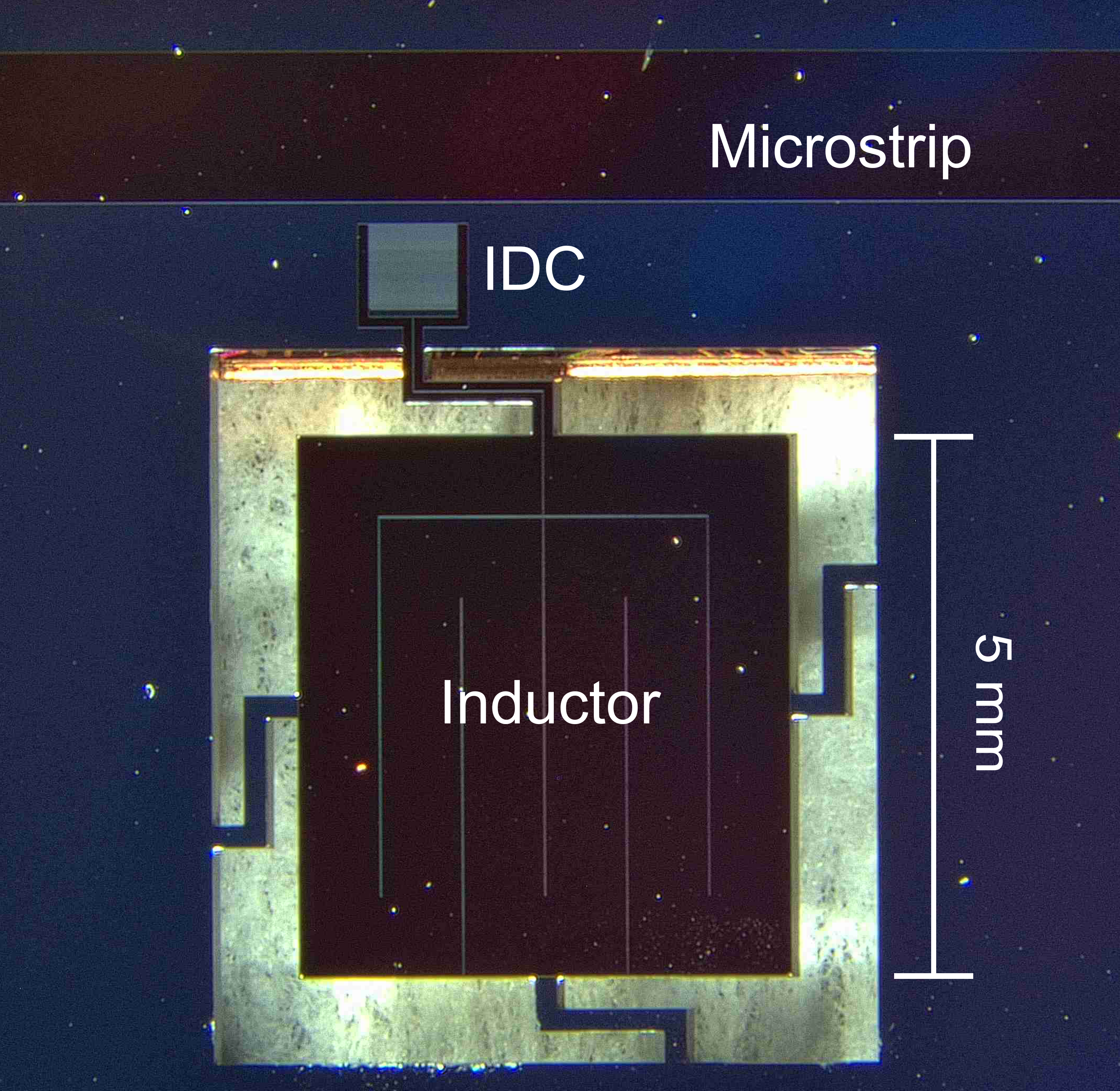}
    \caption{Micrograph of the \qty{500}{\um} thick TKID\@. The TiN-Ti resonator structure consists of a meandered inductor on the square island and an interdigitated capacitor (IDC) on the bulk, ``frame'' substrate. The IDC is capacitively coupled to the microstrip transmission line running horizontally across the top of the image. The \qtyproduct{5 x 5}{mm} island is thermally coupled to the substrate frame only through its four, \RaisingEdge-shaped legs and is thus ``semi-isolated'' from the rest of the substrate. The legs are approximately \qty{200}{\um} wide and \qty{1500}{\um} long. Gold films and gold wire bonds on the outer, frame regions of the chip (not pictured) are used to thermalize the entire substrate.
    } \label{fig:TKID_micrograph}
\end{figure}

We have fabricated a pair of TKID calorimeters to probe radiation backgrounds~\cite{Scott2022}, the main difference between the two devices being the thickness of the silicon substrate (\qty{500}{\um} and \qty{1500}{\um}). The fabrication began with the deposition of a superconducting titanium nitride (TiN) layer. The stoichiometry, and therefore superconducting critical temperature, $T_{C}$, of the film was controlled through the proximity effect~\cite{Werthamer1963, Martinis2000} by depositing alternating layers of Ti and TiN~\cite{Vissers2013}. The 14 TiN films each have nominal thickness \qty{5}{nm}, and the 13 intervening Ti films are each \qty{10}{nm} thick, giving the overall thin film a nominal thickness of \qty{200}{nm}. The $T_{C}$ of this composite superconducting film was measured to be \qty{850}{mK}. The film was lithographically patterned and etched to form the resonator and microstrip transmission line structures. Additionally, gold films were deposited on the outer regions of the chip to support thermalization to its copper enclosure through gold wire bonds. Finally, a deep reactive ion etch (DRIE) was used to define and semi-isolate the radiation-absorbing island. The island is \qty{5}{mm} square for both devices. The deep etch goes entirely through the wafer, and the TKID island thickness is simply equal to the original substrate thickness. This island area was chosen to be representative of a typical quantum circuit. The thinner wafer is in the range commonly used for superconducting circuits, while the thicker one was chosen to help us validate models of scaling with thickness. Figure~\ref{fig:TKID_micrograph} depicts the \qty{500}{\um} thick device. The devices were operated inside a cryostat (see Figure~\ref{fig:cartoon} for context) at a temperature of \qty{175}{mK}, well below the superconducting film's critical temperature.

\section{TKID Measurements of Radiation Backgrounds}
\label{sec:measurements}

The readout of TKIDs and the data analysis procedures are described in detail in the Supplementary Material, Section~\ref{sec:SM_measurements}. In short, a microwave probe tone at the resonant frequency is sent through the cryostat, coupled to the resonator, amplified, and mixed against a reference tone to determine the amplitude and phase of the transmitted tone. Changes in the complex transmission indicate a shift in the resonant frequency due, in this case, to temperature changes in the TKID island. These values are digitally sampled once every \qty{800}{ns}. The TKID response to a radiation event is a transient pulse.



One important requirement for this measurement is sensitivity over a wide range of energies with reasonably good linearity. Changes in the TKID resonant frequency are nearly proportional to changes in the TKID temperature for a wide range of energies deposited in the device, up to several \unit{MeV}. Deposition of energy $E$ heats the TKID island, increasing the temperature $T$ by $\delta T=E/C(T)$, where $C$ is the (weakly temperature-dependent) heat capacity. The equilibrium population of quasiparticles (broken Cooper pairs) in the superconducting film depends on temperature as $n_\mathrm{QP} \propto\sqrt{T} \mathrm{exp}(-\Delta/kT)$~\cite{Day2003}, where $\Delta$ is the superconducting gap energy. The gap $\Delta$ equals $1.76 k T_\mathrm{C}$ in BCS theory~\cite{Tinkham2004}. Although quasiparticle population changes are not linear for large changes in temperature, we have designed the TKID with a high enough heat capacity that even the most energetic events increase $T$ by no more than \qty{1}{\percent}. Changes in quasiparticle density, in turn, cause proportional changes in the thin-film resonator's inductance, and proportional shifts in the resonant frequency. Though this chain of relationships contains multiple departures from strict proportionality, most are small for MeV-scale events in our TKIDs. Furthermore, the positive curvature of the function $n_\mathrm{QP}(T)$ and the fact that silicon's cryogenic heat capacity grows with temperature act as nonlinearities of opposite sign. We estimate that the nonlinear effects due to heat capacity and quasiparticle population change our energy estimates by no more than \qty{20}{\percent} at \qty{10}{MeV} (the highest energy detected in the background spectrum). Below \qty{1}{MeV}, they are less than \qty{3}{\percent} and subdominant to the overall energy-scale calibration uncertainty.

The radiation backgrounds were measured for 168.0 hours with the \qty{500}{\um} TKID and for 97.5 hours with the thicker \qty{1500}{\um} device, both in the horizontal orientation (as defined in Figure~\ref{fig:CR_TKID_loss}). These data were collected with hourly gaps of approximately 100 seconds in which the resonator transmission was re-characterized in case of drifts. In addition to the background data, 5.0 (3.0) hours were devoted to measurements with the thin (thick) TKID of a sealed $^{153}$Gd radiation check source. This source produces emissions of known energies, which we use for energy calibration.

The TKID pulses are found to return to baseline with two exponential time constants, typically around \qty{60}{\micro\s} and \qty{240}{\micro\s}. The latter, slower time constant corresponds to the ratio of the island's heat capacity to the thermal conductance of the TKID island's four legs. To optimize the energy resolution and linearity, we only estimate the amplitude from the slower, thermal component from each pulse and discard information about the (small) amplitude of the faster component.

The TKID shown in Figure~\ref{fig:TKID_micrograph} is sensitive primarily to energy deposited in the square island, but it is not fully immune to the effects of particles that deposit energy in the surrounding frame. Tests were performed with a pulsed diode laser (wavelength \qty{635}{nm}) and carried into the cryostat on an optical fiber. The fiber output was aimed first at the island and then at the frame, showing that optical energy absorbed in the frame can produce a TKID signal as large as a few percent of the signal produced by equivalent optical events aimed at the center of the island. The larger ``frame events'' can be identified and removed from the data because of their unusually slow signal rise and fall times. However, the smaller frame event signals cannot be reliably identified. As a result, the measured flux at inferred energies of less than approximately \qty{200}{keV} is subject to systematic uncertainties and almost certainly overestimates the true flux incident on the TKID island.

The most intense emission from the $^{153}$Gd calibration source is a pair of gamma rays at \qty{97}{keV} and \qty{103}{keV}, but silicon wafers as thin as our TKIDs have negligible photo-absorption at these energies. Instead, the characteristic K$\alpha$ and K$\beta$ emission of the europium decay product was used for calibration. These x-ray lines (\qty{41}{keV} and \qty{47}{keV})~\cite{deslattes2003} are unresolved by the TKID and produce a weighted-average energy of \qty{42.2}{keV}. The energy calibration is anchored to this blended emission line. To confirm that the calibration line was indeed x-ray emission of the excited Eu and not the \qty{100}{keV} gamma rays, we verified that the observed line intensity was reduced by the expected factor of 2.2 when a copper foil \qty{203}{\micro\m} thick was placed in between the check source and the sensor, rather than the factor of 1.1 that would be expected if the observed peak were actually \qty{100}{keV}~\cite{xcom}.


The \qty{42}{keV} calibration peak also establishes an approximate energy resolution of \qty{10}{keV} (full-width at half-maximum) in the thicker TKID and \qty{20}{keV} in the thinner device. This resolution is far from the intrinsic thermal-fluctuation limit of the TKID calorimeters. Possible reasons for the poor resolution include a TKID response that depends on where specifically energy is deposited within the TKID island, non-optimal microwave readout tone power or frequency, and gain drift. It is uncertain whether the resolution is a constant energy value, or a fixed fraction of the deposited energy, or something in between.

Gain stability is difficult to assess from such a spectrum. The $^{153}$Gd check source shows the \qty{42}{keV} peak changing by no more than $\pm$\qty{10}{\percent} from one 30-minute measurement to the next. The pulsed laser coupled to the TKID via an optical fiber shows evidence of gain variations at least this size over ten-minute timescales. Gain drifts do not seem to limit the energy resolution of the current device, yet are large enough to require improvement in future systems. It is not clear whether the drifts are due to changes in the cryostat temperature, in the magnetic environment, or other factors.

The same laser was used with a variable optical attenuator to check the linearity of the TKID sensor. The sensor response is consistent with linearity. The data show a response that grows with the optical pulse energy as $\propto E^\alpha$ for $\alpha=1.03\pm0.05$. This value of $\alpha\approx 1$ is consistent with linearity up to pulses approximately 100 times the amplitude of the \qty{42}{keV} events from the check source. This verifies the energy response up to \qty{4}{MeV} and shows it to be both close enough to linear for the assessment of radiation backgrounds, and consistent with our theoretical understanding of nonlinear effects below \qty{10}{MeV}. This linearity test is itself limited by our uncertainty in how stable the TKID-plus-laser is as a consistent energy standard.

The limitations on TKIDs' energy resolution and linearity and on our ability to perform energy calibrations will be addressed in future sensor designs and future readout optimizations. Thus, although the TKID is a new technology among superconducting microcalorimeters that will benefit from further study and optimization, we will nonetheless demonstrate it is a valuable tool for measuring energy deposited in a superconducting circuit by charged-particle and gamma-ray backgrounds.

\section{Comparison of Model and Measurement}
\label{sec:comparison}

\begin{figure*}
    \centering
    \includegraphics[width=\textwidth]{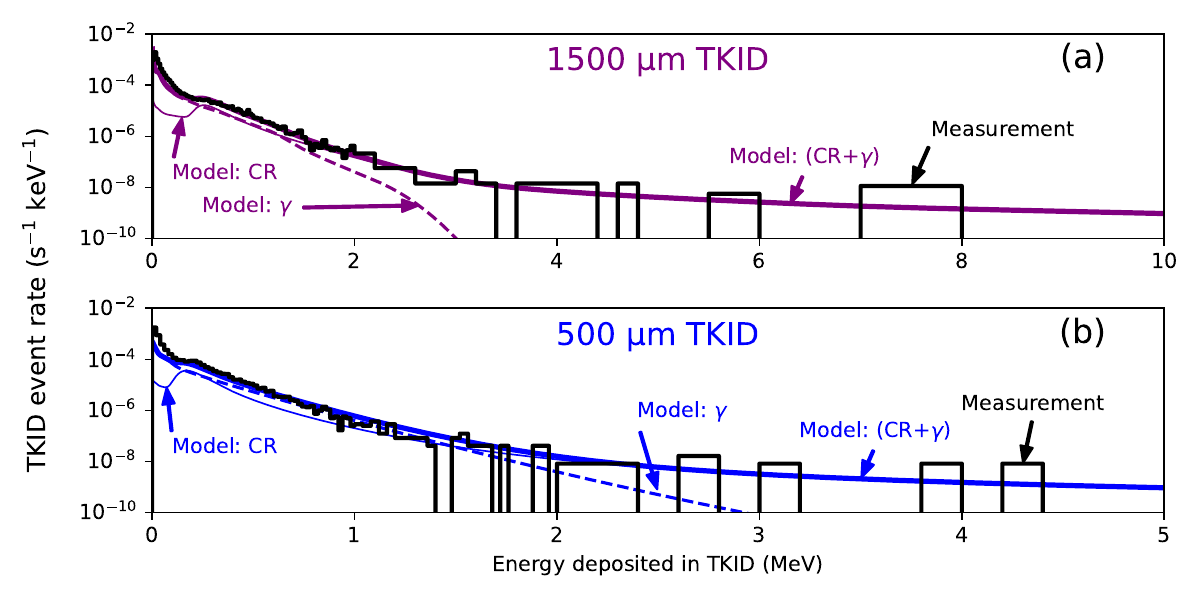}
    \caption{Measured background spectrum from TKIDs made of silicon \qty{1500}{\micro\m} thick (\emph{a}) and \qty{500}{\micro\m} thick (\emph{b}), both in the horizontal orientation (as defined in Figure~\ref{fig:CR_TKID_loss}). Also shown in each panel are the model predictions for terrestrial gamma rays (``Model: $\gamma$'', dashed) and cosmic rays (``Model: CR'', thin solid line), and their sum (thicker solid line). The measured spectrum uses unequal bins to reduce visual distraction where events are rare: there are 50 bins per MeV below \qty{1}{MeV}; 20 per MeV up to \qty{2}{MeV}; 5 per MeV up to \qty{5}{MeV}; and 2 per MeV above \qty{5}{MeV}. The measurements and models agree well, within a factor of 1.5 through most of the energy range studied. The discrepancy at the lowest energies arises because the measurement includes frame events (as defined in the text), while the model curves exclude any frame events. Figure~\ref{fig:results_zoom} shows the low-energy region in more detail.
    }
    \label{fig:results}
\end{figure*}

\begin{figure}
    \centering
    \includegraphics[width=\columnwidth]{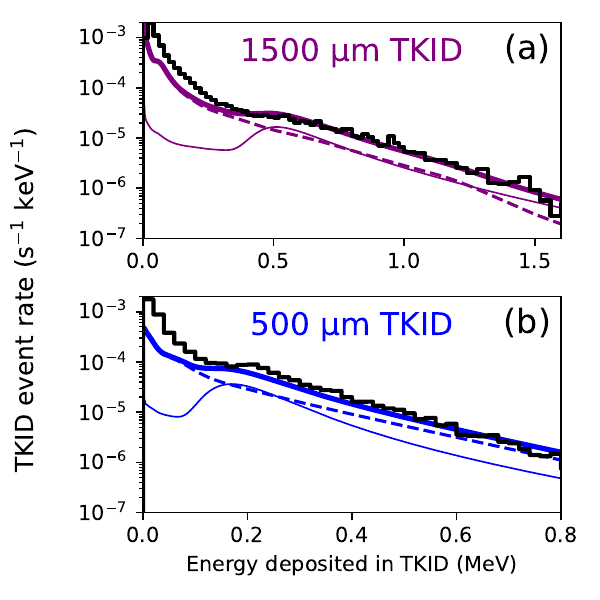}
    \caption{Measured background spectrum from TKIDs made of silicon \qty{1500}{\micro\m} thick (\emph{a}) and \qty{500}{\micro\m} thick (\emph{b}) compared to models of the spectra. The plotted curves are the same as in Figure~\ref{fig:results} but show only the lower-energy events.
    }
    \label{fig:results_zoom}
\end{figure}

\begin{table}[]
    \centering
\begin{tabular}{r@{\ \ }lS[table-format=3.2]S[table-format=3.2]@{\ \ \ \ }S[table-format=3.2]}
$E_\mathrm{min}$ & & {Event} & {Power} & {Mean}\\
(\unit{keV}) & {Component}       & {Rate (\unit{s^{-1}})} & {(\unit{keV.s^{-1}})} & {$E$ (\unit{keV})} \\ \hline
0 & Gamma rays      & 0.0255 & 3.14 & 123 \\
0 & Cosmic rays     & 0.0080 & 1.92 & 241 \\
0 & Model total     & 0.0332 & 5.06 & 153 \\
\\
40 & Gamma rays      & 0.0153(7) & 2.98(15) & 194 \\
40 & Cosmic rays     & 0.0075(2) & 1.91( 6) & 255 \\
40 & Model total     & 0.0227(7) & 4.89(15) & 215 \\
\\
40 & Model $+$ frame   & 0.035\,\ (4) & 5.9(4) & $\cdots$ \\
40 & TKID measured   & 0.0331(2) & 6.3(3) & 190 \\
\end{tabular}
%
    \caption{Model and measurement of the rate of energy-absorption events, the power they deposit, and the mean energy per event for the \qty{500}{\um}-thick silicon substrate. The substrate is square with area \qty{25}{mm^2}. All values represent integrals from $E_\mathrm{min}$ to \qty{20}{MeV}. The first three rows represent the full integral (that is, from a minimum energy of 0), relevant for instruments with high sensitivity to even the smallest background events. The next three rows represent the models integrated starting at \qty{40}{keV}, where the current measurements are most reliable. These rows give the results of the gamma-ray model (Section~\ref{sec:terrestrial}), the cosmic-ray model (Section~\ref{sec:CR}), and their sum. The Model $+$ frame row also adds a model of the excess events that are detected in the TKID island even though the energy was deposited in the frame; this row is most appropriate for comparison to the measured rates. The estimated uncertainties in parentheses are all dominated by various systematic uncertainties, except for the measured TKID event rate, which is dominated by Poisson statistical uncertainty. The frame-hit model has \qty{30}{\percent} relative uncertainties. Equivalent results for the thicker silicon substrate appear in the Supplemental Material Table~\ref{tab:loss_results_thick}.
    }
    \label{tab:loss_results}
\end{table}

Figures~\ref{fig:results} and \ref{fig:results_zoom} show the radiation background spectrum measured by both TKIDs, compared to our background models. This comparison is not a fit to the TKID data---the models are scaled strictly according the NaI measurements. According to the models, the lowest-energy events are primarily due to terrestrial gamma rays, while events depositing \qty{2}{MeV} or more are mostly the result of cosmic rays (primarily protons and neutrons, as Figure~\ref{fig:CR_TKID_components} shows). At intermediate energies, both sources of backgrounds are of comparable intensity. A peak in the cosmic-ray component of the model appears near \qty{520}{keV} in the \qty{1500}{\micro m} sensor and near \qty{170}{keV} in the \qty{500}{\micro m} sensor (Figure~\ref{fig:results_zoom}). This peak results from the abundance of minimum-ionizing particles ($e$ and $\mu$) traversing similar path lengths through the silicon, paths approximately equal to the substrate thickness. Because the gamma-ray spectrum equals or exceeds the cosmic-ray spectrum at these energies, the predicted cosmic-ray peak is largely obscured. It appears in the measurements (and the composite model) as a subtle feature rather than a clear hump.

The measurement has a systematic uncertainty of at least \qty{\pm 5}{\percent} on the energy scale. The limited number of events at the calibration line of \qty{42}{keV} and the lack of calibration features above that energy are the primary reasons for this uncertainty. The energy-scale systematic alone suffices to make the data and model fully consistent across most of the wide energy range measured here, though we have \emph{not} used this freedom to adjust the measurements depicted in Figures~\ref{fig:results} and \ref{fig:results_zoom}. 

For measured energies below approximately \qty{200}{keV}, an additional source of uncertainty and bias in the event rate is present: particles that deposit energy in the silicon frame can produce an unwanted thermal signal in the TKID island. As discussed in Section~\ref{sec:measurements}, we attempted to identify and eliminate such frame events from the spectrum based on their distinctive pulse shape, but reliable discrimination is difficult. Frame hits cause small pulses with low signal-to-noise ratios. The result is a limited effectiveness of this test for events with apparent energy less than $\sim$\qty{200}{keV}, which produces a bias towards overestimation of the spectrum at lower energies. In the Supplemental Material (Appendix~\ref{sec:additional_figs_tabs}), we discuss a model of frame hits that explains the measured excess.  However, the model curves shown in Figures~\ref{fig:results} and \ref{fig:results_zoom} do not include an estimated frame-hit contribution.

Across the range of energies from \qty{200}{keV} to \qty{10}{MeV}, the measurement of the spectrum agrees well with the model of the background radiation, almost everywhere to within a factor of 1.5 in the count rate. We stress that no free parameters were fit to match the models to the measurement---all re-scalings in the intensities of the models were determined from the entirely independent NaI measurements. Figures~\ref{fig:results} and \ref{fig:results_zoom} present absolute-rate comparisons of model and measurement. This means that our radiation models, their intensities normalized to measurements taken by a NaI spectrometer with a volume of \qty{347000}{mm^3}, also describe the spectrum of energy deposited into silicon structures with volumes as small as \qty{12.5}{mm^3} inside a millikelvin refrigerator. Therefore, we conclude the models represent valid tools to predict the impact of background radiation on millimeter-scale silicon devices such as superconducting quantum circuits.

Table~\ref{tab:loss_results} summarizes the comparison between model and measurement for the \qty{500}{\um} TKID, integrated over energy. (The analogous result for the thicker TKID is in the Supplemental Material, Table~\ref{tab:loss_results_thick}). The base model shows a lower rate of events and of power deposited in the island compared to the measurement. When the model is enhanced by the expected consequences of frame hits, however, the measurements agree well. The predicted event rate is only \qty{7}{\percent} higher than the measurement, while the power is \qty{7}{\percent} lower. We consider this to be very good agreement given the simplified models, and also given that the spectrum was integrated well down into the range where frame events and island events cannot be fully discriminated. The rates quoted in the table apply to an island with thickness \qty{500}{\um} and area of \qty{25}{mm^2}, but to a good approximation, the values for other sizes at fixed thickness should be proportional to the island area. Section~\ref{sec:mitigation} and Figure~\ref{fig:thickness_scaling} explore the scaling to thinner substrates.
 
\section{Reducing the Impact of Background Radiation on Quantum Circuits}
\label{sec:mitigation}

\begin{figure*}
    \centering
    \includegraphics[width=\textwidth]{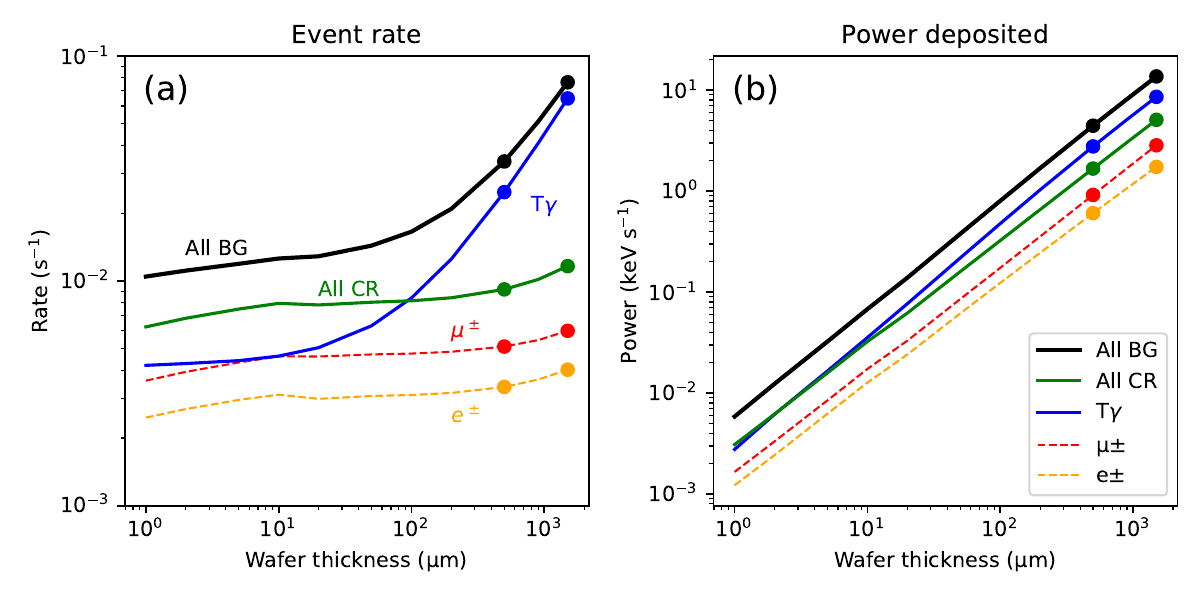}
    \caption{Dependence of event rates (\emph{a}) and background power (\textit{b}) on silicon wafer thickness, integrated from \qty{40}{keV} to \qty{20}{MeV} as in Table~\ref{tab:loss_results}. Rates and power are modeled for a \numproduct{5 x 5}{\,mm$^2$} device inside a simplified model of a cryostat; both should scale approximately as the device area. The values come from the models validated by our TKID measurements, extended to thinner substrates. Thicknesses modeled include \qtylist{1;2;5;10;20;50;100;200;500;900;1500}{\um}. The two thicknesses corresponding to our TKID measurements are indicated by circular markers. Dashed lines are the $\mu^\pm$  and $e^\pm$ components of cosmic rays. Because interactions due to the $p$, $n$, $\gamma$, and $\pi^\pm$ components are far less common, with rates well below \qty{3e-4}{s^{-1}}, they are not shown separately. The sum of all cosmic-ray components is green. The events caused by terrestrial gamma-ray emission (``T$\gamma$'') are shown in blue. The sum of all backgrounds (``All BG'') is black. While the power scales with substrate thickness, cosmic rays impose a minimum event rate for even the thinnest substrates.
    }
    \label{fig:thickness_scaling}
\end{figure*}

Our results shed light on the effectiveness of various strategies for mitigating the effects of background radiation on superconducting circuits, such as TKIDs and qubits. We argued in Section~\ref{sec:CR_qubit} that a chip's orientation (vertical or horizontal) has limited effects. Still, mitigation strategies can be found. Those that we consider include the use of smaller substrate volumes, shielding, and thermal isolation of critical areas; others are possible.

The curves of Figure~\ref{fig:results} show that at any given  energy above \qty{0.5}{MeV}, the rate of background events is roughly a factor of 5--10 higher in the  \qty{1500}{\micro\m} thick device compared to the \qty{500}{\micro\m} device.  After integrating the predictions of our background model over energy, we find that the total event (power) rate is 2.3 (2.8) times higher in the \qty{1500}{\micro\m} device than in the \qty{500}{\micro\m} device.  Clearly, superconducting circuits on smaller substrates will experience less disruption from background radiation.  A substrate can be made smaller by reducing either its area or its thickness.

It is useful to consider the spectra of background events in substrates spanning a wider range of thickness than the two values probed experimentally in this work.  The typical thickness of a \qty{300}{mm}-diameter silicon wafer is \qty{775}{\micro\m}, so substrates thicker than our \qty{1500}{\micro\m} sensor are unlikely.  However, substrates thinner than \qty{500}{\micro\m} are both realistic and interesting as a means of reducing the impact of naturally occurring radiation backgrounds. Silicon becomes difficult to handle at thicknesses below \qty{100}{\um}, but the device layer in silicon-on-insulator (SOI) wafers can be thinner than \qty{1}{\um}. Standard micro-machining techniques can be used to remove the handle layer from underneath selected circuit regions on SOI substrates. Such techniques offer a path to substrates with thickness of a few \unit{\micro m} or less.  Consequently, in Figure~\ref{fig:thickness_scaling} we show models estimating the event rate and dose rate for substrates with thicknesses from \qty{1}{\micro m} to \qty{1500}{\micro\m}.  

Cosmic-ray event rates and power deposition in \qty{500}{\um} wafers are dominated by charged particles (primarily $\mu^\pm$ and $e^\pm$), as shown in Figure~\ref{fig:CR_TKID_components}. Models over the wider range of wafer thickness (Figure~\ref{fig:thickness_scaling}) display the expected behavior: cosmic-ray power grows in proportion to the wafer thickness, while the event rate stays approximately constant. The event rate does grow slightly with the thickness, as an increasing lateral area of the silicon island is exposed to cosmic rays arriving from non-vertical directions. 

The effects of terrestrial gamma rays depend differently on the wafer thickness. For thicker wafers, the event rate is proportional to thickness, as expected given the low probability of gamma-ray scattering. The rate approaches a non-zero value for thin substrates, however, due to incident gamma rays that eject secondary electrons from nearby packaging.  This minimum suggests that a final shield around the qubit could be beneficial, particularly for thinner wafers. Relative to a higher-$Z$ metal of equal mass, a shield made of a low-$Z$ material (such as graphite) would block electrons more effectively~\cite{ICRU1984} while interacting less with gamma rays, suppressing the rate of events due indirectly to terrestrial gamma rays~\cite{Workman:2022ynf}.

The power deposited by terrestrial gamma rays grows faster than the thickness, because the primary mechanism for interaction of MeV-scale gamma rays with silicon is Compton scattering. Scattered electrons are not only more numerous in thicker wafers, but they also travel longer paths and are able to deposit a larger fraction of their energy.  Overall, the use of thinner substrates reduces the power deposited by background radiation in proportion to the thickness reduction, but the event rate falls no lower than approximately \qty{4d-4}{s^{-1}.mm^{-2}}, even for very thin substrates.

An interesting result shown in Figure~\ref{fig:thickness_scaling} is that muons cause the majority of the disruptive cosmic-ray events and power, yet electrons' contributions are only a factor of two smaller at our elevation. This fact suggests that shielding an instrument from most $e^\pm$ secondaries could appreciably reduce cosmic-ray events and power, without requiring a muon-shielding, deep underground site.

The structure of our TKID devices shows that micro-machining can produce silicon islands connected to a surrounding frame of silicon only by narrow beams or legs. Circuitry placed on such an island is partially thermally isolated (\textit{i.e.}, ``semi-isolated'') from background events that occur in the frame, suggesting the use of micromachined moats to isolate critical areas of a superconducting quantum circuit from the larger surrounding chip. Measurements illuminating the frame area with a fiber carrying \qty{635}{nm} laser light to a specific spot \qty{1}{cm} from our TKID showed a thermal response in the TKID approximately \qty{3}{\percent} of the response from direct illumination of the island, thus demonstrating the viability of such protection schemes.  The effectiveness of a moat will depend heavily on the details of its implementation; superior protection should be achievable with thinner or longer legs than those used here, along with additional heat sinking of the silicon frame.

Sapphire substrates are also used for superconducting qubits~\cite{Place2021,Lisenfeld2019,Thorbeck2023}. Though qubit behavior and stability may differ, cosmic rays and gamma rays transfer energy to sapphire by the same mechanisms we have modeled for silicon and are expected to yield quantitatively similar spectra for equal column densities. The cross sections for Compton scattering of photons and the Bethe-Bloch mean energy-loss formulas for charged particles are nearly equal for sapphire and silicon~\cite{Workman:2022ynf}. Given the densities of \qty{4.0}{g.cm^{-3}} for sapphire and \qty{2.3}{g.cm^{-3}} for silicon, losses and event rates in a \qty{300}{\um} sapphire wafer and in a \qty{500}{\um} silicon wafer should be nearly the same. In a full GEANT4 model of cosmic rays, we find the spectrum of events is nearly identical above \qty{100}{keV} in the two wafers (Supplemental Material Figure~\ref{fig:sapphire_compare}). The thicker silicon wafer shows more events at lower energies, in a ratio of 5/3, as expected given their different lateral areas. GEANT4 models of the terrestrial gamma-ray spectrum also show equivalence between the two substrates with equal column density.

In an unshielded laboratory setting, and for any realistic substrate thickness, gamma rays deposit more power than cosmic rays and cause more---or at least a comparable number of---events (Figure~\ref{fig:thickness_scaling}). To shield superconducting quantum circuits from gamma rays is therefore worthwhile. Attenuation of gamma rays by a factor of $e$ at \qty{1}{MeV} requires only \qty{21}{mm} of steel or \qty{12}{mm} of lead~\cite{xcom}, though one must account for muon-induced secondaries produced in such a shield. In the complete absence of gamma rays, the power we observed would be reduced by a factor of three for the thicker wafers, and the event rate by a factor of at least five. The relative benefits would be smaller but still important for wafers thinner than \qty{100}{\um}.

If gamma rays could be fully screened, the remaining cosmic contribution would be dominated by $\mu^\pm$. Muons are highly penetrating and can only be attenuated by operating underground. Still, the most energetic and therefore the most disruptive events are caused by protons and neutrons. Robust superconducting quantum circuits are likely to benefit from a range of shielding strategies, some intended to suppress the most common events and others targeted at the most energetic events.

\section{Conclusions}
\label{sec:conclusions}


We have modeled two sources of naturally occurring background radiation that are dominant in an unshielded surface laboratory: cosmic rays and radiogenic gamma rays from terrestrial sources. We used a commercial NaI detector to measure the intensities of the key components to quantitatively normalize these models. We then checked the validity of these same models for silicon substrates with sizes typical of those used in superconducting qubits. Specifically, we performed spectroscopic measurements of the backgrounds inside a millikelvin cryostat with two TKID devices, superconducting energy-resolving detectors of photons and charged particles. The measurements provide both event rates and incident power levels for silicon substrates that are \qty{500}{\um} and \qty{1500}{\um} thick (Tables \ref{tab:loss_results} and \ref{tab:loss_results_thick}, respectively).  No additional free parameters were required to match model and measurement to within $\pm$\qty{7}{\percent}. For the 500\,(1500)\,\unit{\um} thick device, the modeled average event energy is 215\,(293)\,\unit{keV}.  (The measured average event energies are biased towards lower values by the numerous low energy frame events, as described in the report.)   However, the background events have a broad spectrum whose high energy tail extends well into the MeV range.  Circuit features designed to mitigate the effects of the average background event may be insufficient for events in this high energy tail.  The modeled and measured radiation spectra reported are highly compatible over an energy range that spans a factor of about 25 and an energy-dependent event rate (in units of \unit{s^{-1}.keV^{-1}}) range that spans a factor of about $10^6$.  This success shows that established tools for modeling particle and radiation transport can be used productively to describe the radiation backgrounds in superconducting device substrates, although care is needed in the construction and application of a background radiation model.

Our measurements were made in one specific ground-level laboratory, at moderately high elevation. Still, with appropriate correction factors, the results should apply to any other above-ground facility. Given our location and substrate sizes, we found the effects of gamma rays to be larger than those of cosmic rays, but not overwhelmingly so. The highest energy, MeV-scale events, are entirely the result of cosmic rays, specifically protons and neutrons. Cosmic-ray secondaries would be reduced at sea level, and further for instruments placed only a few meters underground. The effects of gamma radiation can be reduced by the use of low-activity building materials, or by shielding with a few centimeters of steel or lead.  

While these measures are feasible, the difficulty of shielding both terrestrial gamma rays and cosmic rays also argues for the development of superconducting circuit designs that are intrinsically robust to background radiation. We have shown here that the effects of background radiation can be reduced by shrinking the circuit substrate in area or thickness, or by using micromachining to thermally isolate critical substrate regions from the larger chip.  Other mitigation measures are also possible, such as the use of thin metal films to absorb and thermalize phonons created by background events~\cite{Henriques2019,Iaia2022}, superconductor gap-engineering to reduce the quasi-particle density in sensitive circuit regions~\cite{McEwen2024}, or on-chip sensors to detect and veto background events.

Because the TKIDs demonstrated here are compatible with both the fabrication of superconducting qubits and the techniques used to read them out, it should be possible to include TKIDs or closely related MKIDs in quantum circuits. In dispersive qubit readout, the state of the qubit is encoded in the characteristic frequency of another structure, typically a microresonator, that is coupled to a microwave transmission line~\cite{DispersiveReadout}. In both MKIDs and qubits, the resonant frequency is probed with a microwave tone.  In fact, MKIDs and qubits with different frequencies could share a common transmission line. In the future, MKIDs that are co-fabricated with a quantum processor or that are located nearby can act as explicit and unambiguous detectors of background events, offering potential advantages for extremely local active vetoes of background events~\cite{Orrell2021}.

We find several lessons in the modeling we have performed and in the measurements we have made with conventional spectrometers and the TKID superconducting sensors. We find the PARMA model of atmospheric secondary cosmic rays to be good, though the best match to our measurements required rescaling the hadronic component (including muons) by 0.88 and the electromagnetic component by a location-varying factor ranging between 0.5 and 0.8. Apart from the intensity scaling, we argue it is important to include species other than muons in the cosmic ray model: electrons and gamma rays contribute to the backgrounds in silicon at the same energies as muons do, especially at higher-elevation sites. Critically, protons and neutrons generate rare events at substantially higher energies. For terrestrial gamma rays, the spectra acquired in three widely separated sites varied greatly in intensity (by a factor of 4), but not in shape. An assumption of secular equilibrium in the \Uranium\ and \Thorium\ decay chains is supported by the measured spectra, considerably simplifying construction of the background models. The data support a break in equilibrium at the radon step in the \Thorium\ chain with a 3/2 activity ratio. The data do not constrain the pre-radon section of the \Uranium\ chain, due the paucity of energetic gamma rays from that portion of the decay chain. To model the spectrum correctly, it is necessary to account for Compton scattering of the terrestrial gamma-ray background. Finally, for the conditions studied here (an unshielded, ground-level laboratory, with devices on silicon substrates and most reliable for events with $E>$ \qty{40}{keV}), shielding by the cryostat body and the laboratory ceiling plays only a modest role.  As a result, highly simplified geometric models of these components were adequate.

In summary, we present the first spectroscopic measurement of background events in a silicon die whose size and temperature is representative of the substrates used in superconducting quantum circuits. A model of these radiation backgrounds as the result of naturally occurring radioisotopes and of cosmic rays was matched to our specific local conditions via measurements performed with a standard, commercially available gamma-ray spectrometer. This model, without free parameters, agrees quantitatively with spectral measurements of the radiation background when performed with energy sensors similar in size and composition to many quantum circuits. We use the models to demonstrate how different components of the background respond to changes in the substrate thickness. We anticipate that this modeling approach and the measurement result will be broadly useful to the QIS and superconducting sensor communities.

\vspace{1em}

\textbf{Data and code availability: }
The Microcalorimeter Analysis Software System (MASS) repository used for the analysis of the TKID data in this work is available at \url{https://github.com/usnistgov/mass}. The data and all other code used in this study are available from the corresponding author upon reasonable request.

Official contribution of the National Institute of Standards and Technology; not subject to copyright in the United States.

\begin{acknowledgments}
We thank Adam Sirois and Dan Becker for helpful suggestions on this manuscript.
We gratefully acknowledge support from the U.S. Department of Energy (DOE).  This work was supported by the DOE Office of Science, Office of Nuclear Physics, under Award Numbers DE-SC0021415 and DE-SC0023682. Pacific Northwest National Laboratory is a multi-program national laboratory operated for the U.S. DOE by Battelle Memorial Institute under Contract No. DE-AC05-76RL01830.
\end{acknowledgments}


\bibliography{TKID_background}

\appendix
\counterwithin{figure}{section}
\counterwithin{table}{section}



\section{Supplemental Material}
\label{sec:SM}

\subsection{Terrestrial Radiation and Measurements}
\label{sec:NaI_terrestrial}


Some sources of background radiation are sub-dominant in an unshielded, ground-level laboratory and need not be modeled.  Events caused by interactions of decay products from most radio-impurities in the experimental setup (e.g., the dilution refrigerator and the qubit packaging) are expected to be much less frequent than events from external sources~\cite{loer2024}. Potential exceptions are alpha emission from materials with a direct line-of-sight to the surface of the cryogenic circuit, which could contribute significantly.  We have alpha-counted those materials with the largest line-of-sight solid angle to the circuit: the copper and aluminum lids of the housing, and the substrate itself. The measured rates were below the alpha counter’s intrinsic background, corresponding to levels well below the rates of external sources.  Although the radiopurity of the experimental materials may ultimately be important for highly shielded quantum circuits, we concluded that it is not a critical consideration for the present measurement. There are also external sources of alpha and beta radiation (along with gamma rays), but we also ignore these charged-particle backgrounds on the basis that metallic packaging and a cryostat's vacuum shells effectively shield the cryogenic circuit from such radiation. Secondary particles or photons produced by interactions of higher energy particles or photons may be a source of radiation backgrounds~\cite{Du2022}, but these secondaries are already considered in our models of terrestrial and cosmic radiation.

\begin{figure}
    \includegraphics[width=\columnwidth]{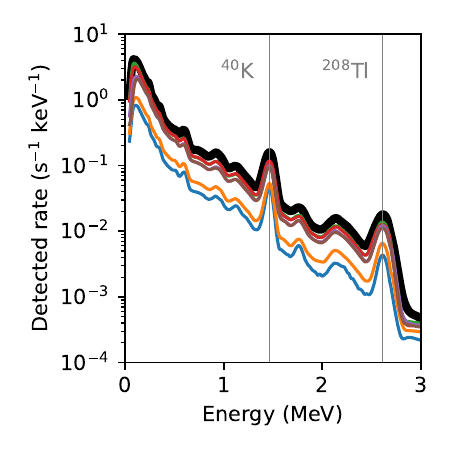}
    \caption{Gamma-ray background recorded by NaI scintillator spectrometer in six different labs at three buildings in two U.S.\ states. The spectrum with the thicker line, our reference spectrum, was measured in the same lab as the resonator measurements described in Section~\ref{sec:TKIDs}. The two energy-calibration features are indicated: \qty{1.461}{MeV} gamma rays from \Potassium\ and \qty{2.615}{MeV} gamma rays from $^{208}$Tl (in the \Thorium\ chain). Terrestrial gamma rays dominate the spectrum in the NaI detector below \qty{2.7}{MeV}, while cosmic rays dominate at higher energies. Variations in the overall intensity of terrestrial gamma rays were the only notable difference among the seven spectra. The two spectra with the lowest gamma-ray activities are those measured in newer buildings.
    }
    \label{fig:NaI-gammas}
\end{figure}

\begin{figure*}
    \centering
    \includegraphics[width=\textwidth]{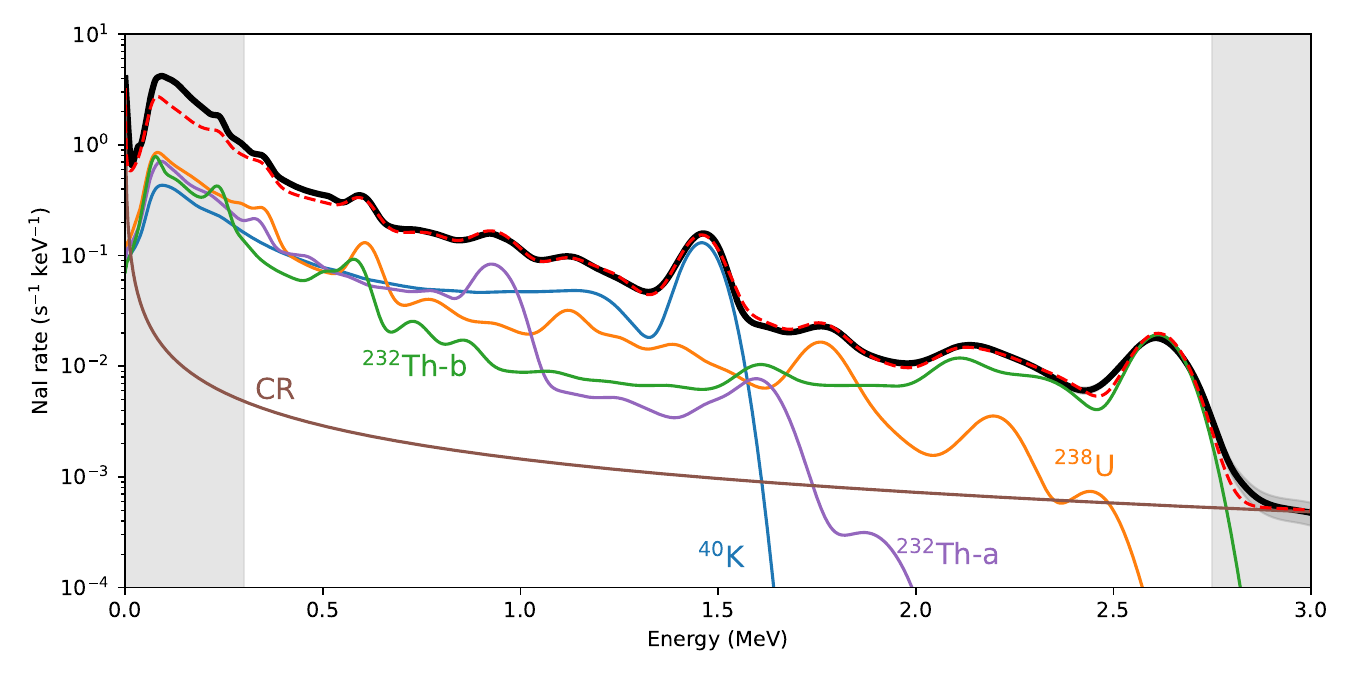}
    \caption{The reference gamma-ray background spectrum measured with the NaI scintillator (solid black) and the model of the same (dashed red). The $y$-axis represents events measured in the NaI spectrometer per unit time, per keV of bin width. The model is the sum of the computed spectrum produced by cosmic rays (CR) plus each of the the three indicated radioactive decay chains (\Potassium, \Thorium, and \Uranium). The Th chain is further divided into pre- and post-radon segments, labeled \Thorium-a and \Thorium-b, which are not assumed to be in secular equilibrium with each other.
    The model was fit to the data between \qtylist{0.3;2.75}{MeV} (i.e., excluding the shaded areas at the left and right of the figure) with the activity of the four decay chain segments allowed to vary. The cosmic-ray level was fixed by measurements at higher energy (Section~\ref{sec:CR}).}
    \label{fig:NaI-fitting}
\end{figure*}

We have estimated the five unknown activities of the terrestrial-gamma model from spectroscopic measurements. For this purpose, we used a commercial gamma-ray detector, consisting of a scintillating NaI crystal mated to a photomultiplier tube. The NaI crystal was a cylinder \qty{76.2}{mm} long and \qty{76.2}{mm} in diameter. The gain scale was calibrated from the two most prominent gamma-ray peaks, those due to \Potassium\ and $^{208}$Tl at \qtylist{1.461;2.615}{MeV}. The measurement was repeated at four laboratories in one building at the National Institute of Standards and Technology (NIST) labs in Boulder, Colorado, with similar results. Two other measurements made in a newer building on the NIST site and at the Pacific Northwest National Laboratory in Richland, Washington showed lower overall intensities but an otherwise similar spectrum (Figure~\ref{fig:NaI-gammas}). At each location, measurements were made in both a high-gain mode (for detection of terrestrial gamma rays) and a low-gain mode (for cosmic rays). The highest detected energies in these modes were approximately \qtylist{4;75}{MeV}, respectively.

The high-gain spectra are described well by the combination of lines and continuum gamma rays from the dominant decay chains. The five activities were fit under the assumption that the NaI spectrometer has a purely Gaussian energy-resolving function with resolution scaling with energy $E$ as $\sigma(E)=A\sqrt{E}$ for some scale factor $A$. A small contribution from cosmic rays was estimated based on the low-gain measurements (Section~\ref{sec:CR}) and fixed during the fit. The energy calibration was performed on the measured spectrum before fitting for activities, so the free parameters of the fit are the five activities (one for each decay-chain segment) and the scale  of the energy resolution, $A$. The best-fit $A$ corresponds to \qty{100}{keV} resolution (FWHM) at \qty{1000}{keV}. Uncertainties in the measured spectrum are assumed to be dominated by Poisson (counting) statistics.

The data prove to be insensitive to the activity of the pre-radon segment of the \Uranium\ chain, which generates relatively few gamma rays. Therefore, we make the simplifying assumption that the entire \Uranium\ chain is in secular equilibrium and repeat the fit with only four unknown activities. Figure~\ref{fig:NaI-fitting} compares the best-fit model and its components with the reference measurement. The agreement is good, though not consistent with purely statistical deviations: the model under-predicts the measured flux below \qty{0.5}{MeV} and seems to underestimate the energy resolution at the $^{208}$Tl peak (\qty{2.615}{MeV}). The statistical uncertainties are approximately \qty{1}{\percent} of the activity values (or \qty{0.3}{\percent} for the \Thorium-b chain). The systematic dependence on details of the fitting procedure is larger, $\sim$\qty{5}{\percent}.

\begin{table}[]
    \centering
    \begin{tabular}{lrrr}
         Isotope & \makecell{Specific activity \\ (\unit{Bq.kg^{-1}})} & 
         \makecell{$\gamma$ escape rate \\ (\unit{cm^{-2}.s^{-1}})}\\ \hline
         \Potassium & 1030 & 1.5 \\
         \Uranium\ chain & 76 & 1.8\\
         \Thorium-a chain & 126 & 1.6 \\
         \Thorium-b chain & 82 & 1.4 \\ 
    \end{tabular}
    \caption{Specific activities determined from NaI scintillator measurements in the reference laboratory. For decay chains, the specific activity is that of the parent isotope; its progeny are assumed to be in secular equilibrium. The \Thorium-a and \Thorium-b activity values refer to the pre-radon and post-radon segments of the decay chain. Emitters are assumed to be uniformly distributed through a concrete layer \qty{22}{cm} thick. The $\gamma$ escape rate is the rate per unit area of photons escaping the upper surface of the floor, after substantial attenuation in the concrete slab (approximately 20 gamma rays escape the top surface for every 100 emitted in the slab). The values are subject to uncertainties of approximately \qty{1}{\percent} (statistical) and \qty{5}{\percent} (systematic).
    }
    \label{tab:terrestrial_activity}
\end{table}

The NaI-based reference measurement allows us to determine the absolute activity of the three decay chains (Table~\ref{tab:terrestrial_activity} for the reference laboratory). Significant intensity variation between locations is expected and observed, but they share the same essential spectral shape. Measurements made in different buildings showed lower absolute activities than our reference measurement, by factors of 2.5 and 4. Widely varying radio-isotope activity levels are the natural result of construction with different concrete and flooring materials. The activity levels in the reference lab are marginally higher than typical for concrete samples but well within the range of natural activities observed in building materials~\cite{Papastefanou2005, Suzuki2000}.

Figure~\ref{fig:NaI-fitting} is very similar to Figure~3(a) of \cite{Vepsalainen2020}. Both show a spectrum of terrestrial gamma rays and model it as a sum of a few decay chains each in secular equilibrium. One difference is that our model explicitly includes a term for the cosmic-ray contribution. Otherwise, the similarity helps to stress the point also shown in Figure~\ref{fig:NaI-gammas}: that the gamma-ray background spectrum is very similar in a wide range of laboratory settings. 


\subsection{Further details on cosmic-ray modeling}
\label{sec:CR_details}

As with the terrestrial gamma rays, we use TOPAS and GEANT4 to model the transport and interactions of the cosmic rays with shielding and detectors. We created an intermediate tool, {\tt MuscRat.jl}~\footnote{A Muon Simulation for Cosmic Ray Analysis Tasks, \url{   https://github.com/joefowler/MuscRat.jl}}, to generate millions of random cosmic rays by sampling from the distributions PARMA produced. These cosmic rays' particle species and their position and momentum vectors were stored as phase-space files, an input data format supported by TOPAS\@. 

The GEANT4 particle-transport framework is readily integrated with a different cosmic-ray generator called CRY~\cite{Hagmann2007}. CRY has been found to under-predict the intensity of near-horizontal muons~\cite{Su:2021}, so we used the CRY+GEANT4 combination only as a check on our models. We did not ultimately find any meaningful differences between the predictions when cosmic-ray secondaries were generated by CRY versus PARMA\@. We preferred PARMA as our cosmic-ray generator mainly because it predicts distributions at arbitrary elevation or atmospheric depth, while CRY works at only three specific elevations.

\begin{figure}
    \includegraphics[width=\columnwidth]{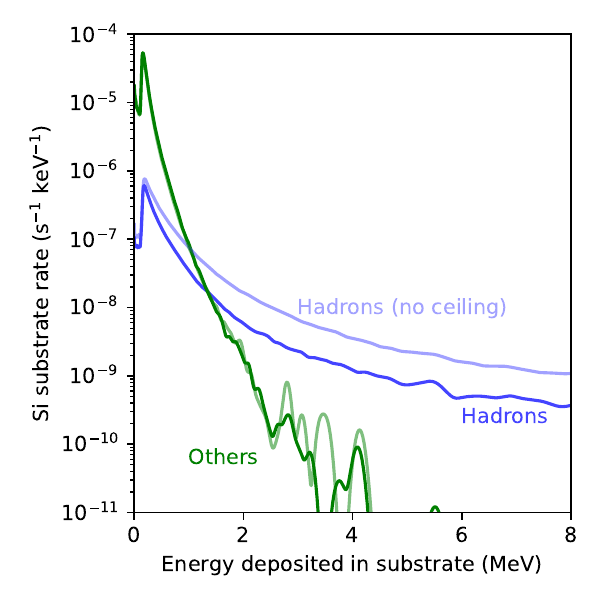}
    \caption{Energy deposited by cosmic-ray secondary particles into a \qty{500}{\um}-thick silicon substrate (models), separated into hadronic particles ($p$, $n$, and $\pi^\pm$) and all others ($\mu^\pm$, $e^\pm$, and $\gamma$). The heavier curves show results for the complete laboratory model, which includes a ceiling (concrete, \qty{18}{cm} thick) and a proxy for the cryostat shell (aluminum, \qty{1}{cm} thick). The lighter curves show the same model without ceiling or cryostat. The ceiling and shell reduce the hadronic spectrum by a factor of approximately 3 for energies $E>$\qty{2}{MeV}, mainly by reducing the flux of lower-energy protons. Protons with sub-GeV energies deposit more energy per distance traveled in the silicon substrate than relativistic protons do and are primarily responsible for the high-energy end of this spectrum~\cite{Workman:2022ynf}. The ceiling and shell reduce the spectrum of energy deposited by all other cosmic-ray particles by no more than a few percent. The analogous distributions for the thicker substrate (not shown) exhibit the same effects.
    }
    \label{fig:CR_TKID_ceiling_effect}
\end{figure}

Our cosmic-ray models include the shielding effect of a concrete ceiling \qty{18}{cm} thick (a value found in building blueprints), as well as the cryostat (here approximated by a simple planar layer of aluminum \qty{1}{cm} thick). These materials reduce the energy carried by muons by only \qty{2.4}{\percent}. They convert a portion of the energy carried by gamma rays into energetic electrons and positrons. The ceiling also reduces the energy carried by neutrons and protons, by approximately \qty{30}{\percent}. 
Because it preferentially attenuates the nucleons of lower energy, which deposit energy more efficiently in a thin object like a silicon wafer, the ceiling and cryostat reduce the rate of rare events in the high-energy end of the spectrum expected in a \qty{500}{\um}-thick silicon substrate by a factor of approximately 3 (Figure~\ref{fig:CR_TKID_ceiling_effect}).  Overall, however, in an above-ground laboratory, the ceiling and cryostat body produce only a small reduction in the cosmic-ray rate and a modest redistribution of energy among particle species.

\subsection{Cosmic-ray measurements with a scintillating-crystal spectrometer}
\label{sec:CR_NaI}

To test the cosmic-ray models, we measured the background spectrum outside the cryostat with the same NaI scintillator described in Section~\ref{sec:NaI_terrestrial}. When configured in a low-gain mode, the spectrometer could measure energetic events up to approximately \qty{75}{MeV}. This energy range is well matched to the energy loss of minimum-ionizing particles in NaI along the possible geometric paths through the cylindrical NaI crystal with length and diameter both \qty{76.2}{mm}. The typical path produces a broad energy-loss peak centered between \qtylist{30;40}{MeV}. As with the measurements of terrestrial gamma rays, we repeated the cosmic-ray measurements in multiple buildings at two sites. One site was near sea level, while the Boulder, Colorado measurements were performed at \qty{1650}{m} above sea level (mean atmospheric depth \qty{860}{g.cm^{-2}}). The observed cosmic-ray flux is approximately \qty{33}{\percent} higher at the higher elevation, consistent with predictions of the PARMA model.

\begin{figure}
    \centering
    \includegraphics[width=\columnwidth]{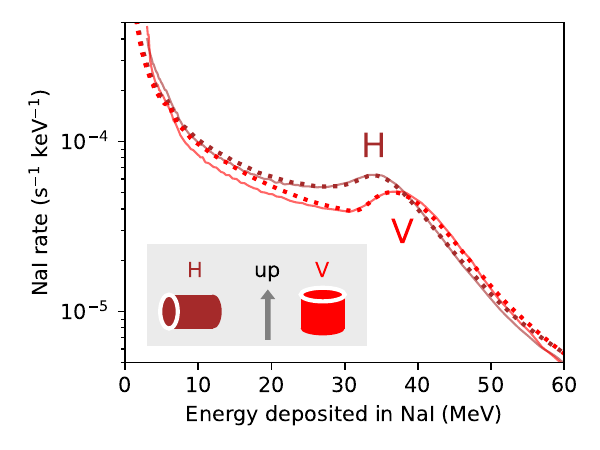}
    \caption{
    Cosmic-ray spectra measured by a NaI scintillator in neighboring labs (solid lines), one with the crystal cylinder oriented vertically (V), the other horizontally (H). The gray inset indicates the orientation conventions. TOPAS+GEANT4 models of the detector in each orientation are shown as dotted lines. The models are scaled in intensity by a factor of 0.88 relative to the PARMA output to agree with the measurement. The energy scale is fixed by calibration to terrestrial gamma-ray features at \qtylist{1.4;2.7}{MeV} then fit to permit a slightly nonlinear response by pinning the peaks at the expected energies of \qty{34}{MeV} and \qty{37}{MeV}. The nonlinearity is found to be a \qty{1}{\percent} effect at these higher energies.
    The spectrum below \qty{3}{MeV} is dominated by terrestrial sources of gamma rays and is not shown here (see Figure~\ref{fig:NaI-fitting}).
    }
    \label{fig:CR_spectra_HV}
\end{figure}

We observed a clear difference between the energy spectrum of cosmic-ray events depending upon whether the NaI cylinder's axis was oriented vertically or horizontally. Because cosmic rays are not isotropic, the distribution of geometric paths through a cylinder is different in the two orientations, and the most probable path length through it is some $\approx$\qty{10}{\percent} longer when vertically oriented. Measurements and model predictions exhibit an identical dependence on cylinder orientation (Figure~\ref{fig:CR_spectra_HV}). The most probable energy loss is \qty{34}{MeV} for a horizontal cylinder and \qty{37}{MeV} for a vertical one. The mean count rate and rate of energy deposition are equal in the two orientations. This observation supports the validity of the particle-transport modeling.

The NaI measurements constrain the intensity of the cosmic rays. 
The NaI spectra shown in Figure~\ref{fig:CR_spectra_HV}, as well as the other four measured at the Boulder site, all agree that the cosmic-ray rate was somewhat lower than predicted by PARMA. To achieve a good fit of the model to the measurements, it was necessary to scale the $\mu^\pm$ intensity separately from the other electromagnetic components ($e^\pm$ and $\gamma$). The hadronic particles (protons, neutrons, and $\pi^\pm$) contribute little to the NaI spectra and are not constrained by our measurements. For simplicity, and given their shared origin in the nuclear component of air showers~\cite{Fowler2001}, we scale the hadronic intensity by the same value as the $\mu^\pm$ component.

The best-fit scaling for our reference spectrum is 0.79 times the PARMA predictions for the electromagnetic shower components ($e^\pm$ and $\gamma$), and 0.88 for the muons and all other species. Measurements made in other rooms on the same site yield scale factors for the PARMA predictions between 0.5 and 0.8 for the electromagnetic components and between 0.85 and 0.88 for the other species. We attribute the variation to the range of shielding overburden. Our cosmic-ray model includes the shielding effect of a particular concrete ceiling \qty{18}{cm} thick. While this model accurately describes conditions of the reference measurement, other spectra were recorded in rooms with additional building levels and more structural materials above them. The NaI measurements show that this additional overburden reduces the $e^\pm$ and $\gamma$ intensity by as much as \qty{40}{\percent} but attenuates the highly penetrating $\mu^\pm$ component by only a few percent. These effects are consistent with the predictions of GEANT4 models for attenuation in the concrete ceiling of the reference measurement.
The rescaling of the PARMA model's $\mu$ intensity by a factor of 0.88 is both small and empirically necessary. We do not know whether this modest disagreement in the cosmic-ray intensity reflects over-prediction by PARMA, or additional screening in the ceiling that we have failed to capture in our models. We use the rescaling factors found from NaI measurements to correct the intensities for all cosmic-ray models when we compare them to measurements made in a superconducting circuit.

\subsection{TKID Measurements of Radiation Backgrounds}
\label{sec:SM_measurements}

Here we describe some details of the TKID operation and readout system that we omitted from the main text.

A standard homodyne mixer is used to collect data with the TKID devices~\cite{Gao2008}. A microwave synthesizer generates a probe tone at the TKID resonant frequency (approximately \qty{1.3}{GHz} for both of our devices). This signal is split, with one portion used as the reference input of an IQ (in-phase and quadrature) mixer and the other portion sent through the cryostat and TKID. This signal is amplified both cryogenically and at room temperature before reaching the RF signal input of the IQ mixer. The $I$ and $Q$ outputs are digitized with a sampling rate of $1.25\times 10^6$ samples per second. Both raw data streams are stored to disk for offline analysis. 

A sweep of probe frequencies near resonance allows us to find the arc in the complex plane described by the transmission of the resonator. Characterizing all samples of the TKID measurement by their angle $\theta$ in the complex plane with respect to the center of this arc, we can replace the $I$ and $Q$ timestreams by a single angle. This procedure allows for more linear response and higher signal-to-noise ratio (SNR) at higher energies. Radiation events are identified in the phase-angle timestream through a level trigger: a pulse is recorded whenever four consecutive samples cross a threshold defined by three times the root-mean-square noise level.

Distinct pulse records were first analyzed by the method of ``optimal filtering''~\cite{Szymkowiak1993}, in which all pulses are presumed to have identical shapes and differ only in amplitude. The method takes account of the non-white nature of the intrinsic noise, and it is explicitly insensitive to slow variations in the DC level of pulse records. 

In violation of the assumptions underlying optimal filtering, however, the TKID pulses exhibit a range of shapes. Specifically, after a rise time of a few \unit{\micro\s} (governed by the resonator bandwidth), pulses are found to fall with two distinct exponential time constants, typically around \qty{60}{\micro\s} and at least \qty{240}{\micro\s}. The slower time constant is consistent with conductive cooling through the TKID island's four legs after the island has thermally equilibrated; it equals the ratio of the island's heat capacity to the thermal conductance of the legs. The cause of the initial, faster decay time is uncertain; it may correspond to a period when the components of the TKID island are out of thermal equilibrium and/or involve electrothermal feedback from the readout tone~\cite{Agrawal2021}. The variation in pulse shape is primarily due to pulses having these two components in a ratio that varies, especially with pulse amplitude. To optimize the energy resolution and linearity by extracting only the amplitude of the slower, thermal component from each pulse, we modified the usual optimal-filter method to fit pulses as sums of the two relevant shapes~\cite{Fowler2017}.

\subsection{Additional Figures and Tables}
\label{sec:additional_figs_tabs}

\begin{figure}
    \includegraphics[width=\columnwidth]{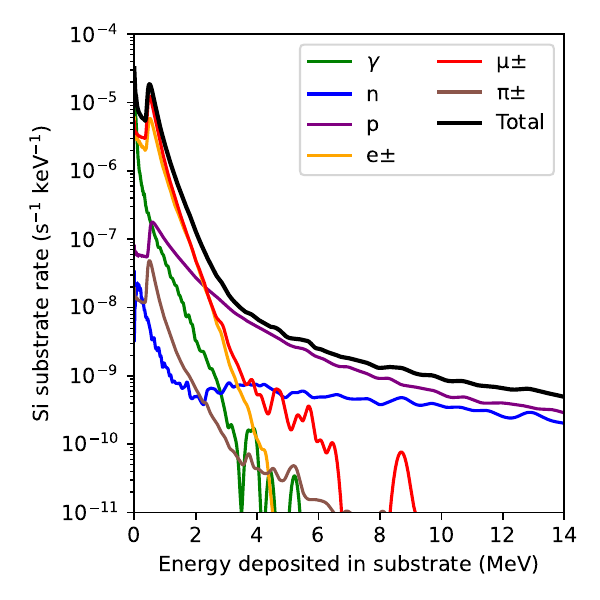}
    \caption{Energy deposited in a \qty{1500}{\um}-thick silicon substrate by cosmic rays (models), both the total and separated by particle species. Similar to Figure~\ref{fig:CR_TKID_components}, which shows the same spectrum components for the thinner silicon substrate (\qty{500}{\um} thick).
    }
    \label{fig:CR_TKID_thick_components}
\end{figure}

\begin{table}
    \centering
\begin{tabular}{r@{\ \ }lS[table-format=3.2]S[table-format=3.2]@{\ \ \ \ }S[table-format=3.2]}
$E_\mathrm{min}$ & & {Event} & {Power} & {Mean}\\
(\unit{keV}) & {Component}       & {Rate (\unit{s^{-1}})} & {(\unit{keV.s^{-1}})} & {$E$ (\unit{keV})} \\ \hline
0 & Gamma rays      & 0.066 & 8.8 & 132 \\
0 & Cosmic rays     & 0.0099 & 5.3 & 530 \\
0 & Model total     & 0.075 & 14.0 & 186 \\
\\
40 & Gamma rays      & 0.038 (2) & 8.3(4) & 221 \\
40 & Cosmic rays     & 0.0090(3) & 5.3(1) & 583 \\
40 & Model total     & 0.046 (2) & 13.6(4) & 293 \\
\\
40 & Model $+$ frame   & 0.087(13) & 17.4(13) & $\cdots$ \\
40 & TKID measured   & 0.0778(4) & 17.4(9) & 223 \\
\end{tabular}
    \caption{Model and measurement of the rate of energy-absorption events, the power they deposit, and the mean energy per event, for the \qty{1500}{\um} silicon substrate. All values represent integrals from $E_\mathrm{min}$ to \qty{20}{MeV}. The first three rows represent the full integral (that is, from a minimum energy of 0), relevant for instruments with high sensitivity to even the smallest background events. The next three rows represent the models integrated starting at \qty{40}{keV}, where the current measurements are most reliable. These rows give the results of the gamma-ray model (Section~\ref{sec:terrestrial}), the cosmic-ray model (Section~\ref{sec:CR}), and their sum. The Model~$+$~Frame row also adds an approximate model of the excess events that are detected in the TKID island even though the energy was deposited in the frame; this row is most appropriate for comparison to the measured rates. Uncertainties are indicated in parentheses and are dominated by systematic effects, apart from the measured TKID event rate, which is dominated by counting statistics. Equivalent results for the thinner device appear in Table~\ref{tab:loss_results}.
    }
    \label{tab:loss_results_thick}
\end{table}

\begin{figure}
    \includegraphics[width=\columnwidth]{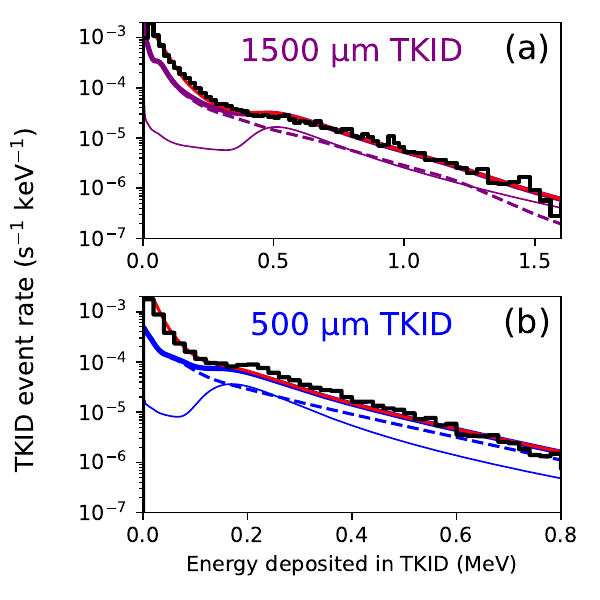}
    \caption{All curves are the same as shown in Figure~\ref{fig:results_zoom}, except for the addition of a red curve. It shows the total model of gamma rays and cosmic rays, \emph{plus} the model of frame hits. The addition of a frame-hit term to the model brings the complete model into good agreement with the measurement (black histogram) below 200\,(400)\,\unit{keV} for the 500\,(1500)\,\unit{\um} substrate.}
    \label{fig:results_zoom_framehits}
\end{figure}

\begin{figure}
    \includegraphics[width=\columnwidth]{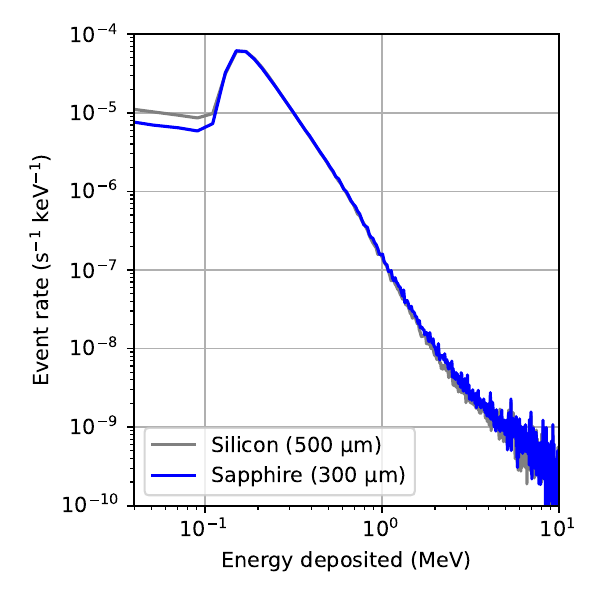}
    \caption{Cosmic-ray models of the spectrum of energy deposited in silicon and sapphire substrates of equal mass: \qty{500}{\um} and \qty{300}{\um} thick, respectively, and each \qtyproduct{5 x5 }{mm}. The spectra differ only at energies $E<$\qty{100}{keV}, where event rates are proportional to the area of the lateral surfaces.}
    \label{fig:sapphire_compare}
\end{figure}

\newpage

We include an additional table and three additional figures as online-only supplemental material. Table~\ref{tab:loss_results_thick} is analogous to Table~\ref{tab:loss_results}, showing energy-integrated event rates and power from models and measurements, but for the case of the TKID fabricated on a \qty{1500}{\um} substrate.

Figure~\ref{fig:CR_TKID_thick_components} shows a model of the distribution of energy deposited in a silicon substrate \qty{1500}{\um} thick by cosmic rays, as well as the breakdown by particle species; it is analogous to Figure~\ref{fig:CR_TKID_components} but for the thicker wafer.

Figure~\ref{fig:results_zoom_framehits} shows that a simple model of power coupling from off-island events can explain the model-measurement discrepancies observed below \qty{350(150)}{keV} in the \qty{1500(500)}{\um} substrate. This coupling is not intended or wanted but must be estimated. In this model, we suppose that the TKIDs are sensitive to energy deposited not only in the \qty{25}{mm^2} TKID island, but also in a further \qty{100}{mm^2} of area from the supporting legs and the nearby regions of frame. The sensitivity in the legs is comparable to that of the island but falls rapidly with distance away from the legs. We have some information that constrains the model. The approximate amplitude of the effect was estimated from the measurements of TKID response to LED illumination of the frame described in the main text. The size scale of the relevant frame area (\qty{100}{mm^2}) and the spatial distribution of the coupling was informed by modeling of Fourier's Law thermal conduction in the island, legs, and frame.  This model, though quantitative, omits ballistic phonons and has large uncertainties. We argue not that it perfectly explains the discrepancy, but that a quantitative model of the thermal coupling between the TKID inductor and the frame with plausible parameters can relieve the tension caused by the clear excess in the measured spectra over models. Future TKID designs will be undertaken with a goal of reducing this sensitivity to energy deposited close to, but not in, the island.

Figure~\ref{fig:sapphire_compare} supports our statements in Section~\ref{sec:mitigation} that cosmic rays passing through two substrates of silicon and sapphire with equal column densities will produce nearly identical spectra, apart from lateral-area effects seen at the low-energy end of the spectra (below \unit{100}{keV}). The figure shows models of cosmic rays only; the terrestrial gamma-ray spectrum computed for the equivalent sapphire substrate is essentially identical to the silicon spectrum shown in Figure~\ref{fig:gamma_qubit} and is therefore not shown.



\end{document}